\documentclass[traditabstract]{aa} % for the abstract without structuration 
\usepackage{natbib}
\usepackage{graphicx}
\usepackage{txfonts}
\usepackage{multirow}
\usepackage{subfig}
\begin{document}
   \title{GOODS-\emph{Herschel}: Ultra-deep \emph{XMM-Newton} observations reveal AGN/star-formation connection\thanks{{\it Herschel} is an ESA space observatory with science instruments provided by European-led Principal Investigator consortia and with important participation from NASA.}\fnmsep\thanks{This work is based on observations obtained with {\it XMM-Newton}, an ESA science mission with instruments and contributions directly funded by ESA Member States and the USA (NASA).}}
%   \titlerunning{GOODS-\emph{Herschel}: Ultra-deep \emph{XMM-Newton} observations reveal AGN/SFR connection}
%   \subtitle{}
   \author{E. Rovilos\inst{1,2}
           \and
           A. Comastri\inst{1}
           \and
           R. Gilli\inst{1}
           \and
           I. Georgantopoulos\inst{1,3}
           \and
           P. Ranalli\inst{1,3}
           \and
           C. Vignali\inst{4,1}
           \and
           E. Lusso\inst{5}
           \and
           N. Cappelluti\inst{1}
           \and
           G. Zamorani\inst{1}
           \and
           D. Elbaz\inst{6}
           \and
           M. Dickinson\inst{7}
           \and
           H. S. Hwang\inst{8}
           \and
           V. Charmandaris\inst{9,10}
           \and
           R. J. Ivison\inst{11,12}
           \and
           A. Merloni\inst{13}
           \and
           E. Daddi\inst{6}
           \and
           F. J. Carrera\inst{14}
           \and
           W. N. Brandt\inst{15}
           \and
           J. R. Mullaney\inst{6,2}
           \and
           D. Scott\inst{16}
           \and
           D. M. Alexander\inst{2}
           \and
           A. Del Moro\inst{2}
           \and
           G. Morrison\inst{17,18}
           \and
           E. J. Murphy\inst{19}
           \and
           B. Altieri\inst{20}
           \and
           H. Aussel\inst{6}
           \and
           H. Dannerbauer\inst{6,21}
           \and
           J. Kartaltepe\inst{7}
           \and
           R. Leiton\inst{6,22}
           \and
           G. Magdis\inst{23}
           \and
           B. Magnelli\inst{13}
           \and
           P. Popesso\inst{13}
           \and
           I. Valtchanov\inst{20}
          }

   \institute{INAF-Osservatorio Astronomico di Bologna, via Ranzani 1, 40127,
              Bologna, Italy
              \and
              Department of Physics, Durham University, South Road, Durham,
              DH1, 3LE, UK
              \and
              Institute of Astronomy \& Astrophysics, National Observatory of
              Athens, Palaia Penteli, 15236, Athens, Greece
              \and
              Dipartimento di Astronomia, Universit\`{a} di Bologna, via
              Ranzani 1, 40127, Bologna, Italy
              \and
              Max Planck Institut f\"{u}r Astronomie, K\"{o}nigstuhl 17,
              D-69117, Heidelberg, Germany
              \and
              Laboratoire AIM, CEA/DSM-CNRS-Universit\'{e} Paris Diderot,
              IRFU/Service d'Astrophysique, B\^{a}t.709, CEA-Saclay, 91191
              Gifsur-Yvette Cedex, France
              \and
              National Optical Astronomy Observatory, 950 North Cherry Avenue,
              Tucson, AZ 85719, USA
              \and
              Smithsonian Astrophysical Observatory, 60 Garden Street,
              Cambridge, MA 02138, USA
              \and
              Department of Physics and Institute of Theoretical \&
              Computational Physics, University of Crete, GR-71003, Heraklion,
              Greece
              \and
              IESL/Foundation for Research \& Technology-Hellas, GR-71110,
              Heraklion, Greece and Chercheur Associ\'{e}, Observatoire de
              Paris, F-75014 Paris, France
              \and
              UK Astronomy Technology Centre, Science and Technology Facilities
              Council, Royal Observatory, Blackford Hill, Edinburgh EH9 3HJ, UK
              \and
              Institute for Astronomy, University of Edinburgh, Blackford Hill,
              Edinburgh EH9 3HJ, UK
              \and
              Max-Planck-Institut f\"{u}r extraterrestrische Physik,
              Giessenbachstra\ss e, 85471, Garching bei M\"{u}nchen, Germany
              \and
              Instituto de F\'{i}sica de Cantabria (CSIC-Universidad de
              Cantabria), Avenida de los Castros, 39005 Santander, Spain
              \and
              Department of Astronomy and Astrophysics, Pennsylvania State
              University, 525 Davey Laboratory, University Park, PA 16802, USA
              \and
              Department of Physics and Astronomy, University of British
              Columbia, Vancouver, BC V6T 1Z1, Canada
              \and
              Institute for Astronomy, University of Hawaii, Manoa, HI 96822,
              USA
              \and
              Canada--France--Hawaii Telescope Corp., Kamuela, HI 96743, USA
              \and
              Observatories of the Carnegie Institution for Science, 813 Santa Barbara
              Street, Pasadena, CA 91101, USA
              \and
              Herschel Science Centre, European Space Astronomy Centre,
              Villanueva de la Ca\~{n}ada, 28691 Madrid, Spain
              \and
              Universit\"{a}t Wien, Institut f\"{u}r Astronomie,
              T\"{u}rkenschanzstra\ss e 17, 1180 Wien, Austria
              \and
              Astronomy Department, Universidad de Concepci\'{o}n,
              Concepci\'{o}n, Chile
              \and
              Department of Physics, University of Oxford, Keble Road, Oxford
              OX1 3RH, UK
             }

%   \date{Received ...; accepted ...}
   \date{Draft version of \today}

\abstract{Models of galaxy evolution assume some connection between the AGN and
               star formation activity in galaxies. We use the multi-wavelength
               information of the CDFS to assess this issue. We select the AGNs from the
               3\,Ms {\it XMM-Newton} survey and measure the star-formation rates of
               their hosts using data that probe rest-frame wavelengths longward of
               $\rm 20\,\mu m$, predominantly from deep
               $\rm 100\,\mu m$ and $\rm 160\,\mu m$ {\it Herschel} observations,
               but also from \emph{Spitzer} MIPS-$\rm 70\,\mu m$. Star-formation rates 
               are obtained from spectral energy distribution fits, identifying and
               subtracting an AGN component. Our sample consists of sources in the
               $z\approx0.5-4$ redshift range, with star-formation rates
               $\rm SFR\approx10^{1}-10^{3}\,M_{\sun}\,yr^{-1}$ and stellar masses
               $M_{\star}\approx10^{10}-10^{11.5}\,{\rm M_{\sun}}$. We divide the
               star-formation rates by the stellar masses of the hosts to derive specific
               star-formation rates (sSFR) and find evidence for a positive correlation
               between the AGN activity (proxied by the X-ray luminosity) and the sSFR for
               the most active systems with X-ray luminosities exceeding
               $L_{\rm x}\simeq10^{43}{\rm \,erg\,s^{-1}}$ and redshifts $z\gtrsim1$.
               We do not find evidence for such a correlation for lower luminosity systems
               or those at lower redshifts, consistent with previous studies. We do not
               find any correlation between the SFR (or the sSFR) and the X-ray absorption
               derived from high-quality {\it XMM-Newton} spectra either, showing that
               the absorption is likely to be linked to the nuclear region rather than the
               host, while the star-formation is not nuclear. Comparing the sSFR of the
               hosts to the characteristic sSFR of star-forming galaxies at the same
               redshift (the so-called ``main sequence'') we find that the AGNs reside
               mostly in main-sequence and starburst hosts, reflecting the AGN - sSFR
               connection; however the infrared selection might bias this result. Limiting
               our analysis to the highest X-ray luminosity AGNs (X-ray QSOs with
               $L_{\rm x}>10^{44}{\rm \,erg\,s^{-1}}$), we find that the highest-redshift
               QSOs (with $z\gtrsim2$) reside predominantly in starburst hosts, with an
               average sSFR more than double that of the ``main sequence'', and we find a
               few cases of QSOs at $z\approx1.5$ with specific star-formation rates
               compatible with the main-sequence, or even in the ``quiescent'' region.
               Finally, we test the reliability of the colour-magnitude diagram (plotting the
               rest-frame optical colours against the stellar mass) in assessing host
               properties, and find a significant correlation between rest-frame colour
               (without any correction for AGN contribution or dust extinction) and sSFR
               excess relative to the ``main sequence'' at a given redshift. This
               means that the most ``starbursty'' objects have the bluest rest-frame
               colours.}
   {}{}{}{}


   \maketitle

\section{Introduction}

One of the most significant observations of modern-day astrophysics is the evidence
that the mass of the super-massive black hole (SMBH) in the centre of any galaxy is
correlated to the properties of its bulge, parametrised by the spheroid luminosity
\citep[e.g.][]{Magorrian1998}, or the spheroid velocity dispersion
\citep[e.g.][]{Ferrarese2000}. This relation has a small intrinsic dispersion
\citep[e.g.][]{Gultekin2009} which implies an evolutionary connection between
the SMBH and the spheroid. The mechanisms that build the super-massive black hole
and the bulge of the galaxy are an active galactic nucleus (AGN) and star-formation
or possibly merging episodes, respectively. There is additional evidence that the
space density of AGNs and cosmic star formation have similar redshift evolution, at
least up to redshifts $z\sim2$ \citep[e.g.][]{Chapman2005,Merloni2008}.

The coeval growth of the SMBH and the host galaxy implies some causal connection
between the AGN and star-formation properties
\citep[see][for a review]{Alexander2012}. Theoretical and semi-analytical models of
galaxy evolution through mergers assume such a connection, where AGN feedback
\citep[e.g.][]{Hopkins2006,DiMatteo2008} plays a catalytic role. After the SMBH has
grown sufficiently massive, the outflows driven by the radiation pressure of the AGN
have enough energy to disrupt the cold gas supply which sustains the star formation
\citep[e.g.][]{Springel2005,King2005}, giving rise to the SMBH-bulge relation. The gas
supply for both the AGN and the star formation is often thought to come from the
galaxy mergers, which are ideal mechanisms for removing angular momentum from
the participant galaxies and funnelling gas to the central kpc region
\citep[e.g.][]{DiMatteo2005,Barnes1996}.

There is, however, growing evidence that a significant part of galaxy evolution takes
place in secularly evolving systems. There is a well-defined relation between the
star-formation rate and the stellar mass in local star-forming systems
\citep[see e.g.][]{Brinchmann2004,Salim2007} which defines the so-called ``main
sequence'' of star formation. This relation is also found in higher redshift galaxies
\citep[e.g.][]{Elbaz2007,Daddi2007} with a redshift-dependent normalisation. It is
also observed that more signs of recent merging activity are found in the
morphology of starbursts (defined as star-forming galaxies with star-formation
rates higher than the main sequence) than in normal (main-sequence) star-forming
galaxies \citep{Kartaltepe2012}. Mapping the star-formation in high-redshift
($z\sim1-3$) galaxies using integral field spectroscopy, \citet{ForsterSchreiber2009}
find that about one third of those star-forming galaxies have rotation-dominated
kinematics showing no signs of mergers. Moreover, \citet{Rodighiero2011} have
shown that $\sim90\%$ of the star-formation density at $z\sim1-3$ takes place in
the main-sequence galaxies. The hosts of AGNs do not seem to significantly deviate
from this main sequence \citep{Mullaney2012,Santini2012}. Similarly,
\citet{Grogin2005} found no apparent connection between mergers and AGN activity
at redshifts $0.4\lesssim z\lesssim1.3$, a result which is also confirmed by
\citet{Cisternas2011} in a similar redshift range ($0.3\lesssim z\lesssim1.0$), and by
\citet{Kocevski2012} at higher redshifts ($1.5\lesssim z\lesssim2.5$). In this case,
gravitational instabilities of the system may cause the transfer of material to the
centre through the formation of bars and pseudo-bulges \citep{Kormendy2004}.
\citet{Hopkins2010a} and \citet{DiamondStanic2012} connected the black-hole
accretion to the nuclear star-formation. In this study, we expand the search for an
AGN-host connection to higher redshifts.

Observationally the identification of a connection between the star-formation and
accretion rates is challenging, especially at high redshifts. The most efficient way is
to isolate the characteristic emission bands of both processes, namely the hard
X-ray emission from the hot corona of the AGN and the far-infrared emission from
cold dust heated by the UV radiation of massive young stars or radio synchrotron
emission from electrons accelerated in supernova explosions. Previous studies using
those indicators in deep fields have shown hints of a correlation
\citep[e.g.][]{Trichas2009}, which is more prominent in AGNs with higher luminosities
and redshifts \citep{Mullaney2010,Lutz2010,Shao2010}. These results argue in
favour of different mechanisms, secular evolution and evolution through mergers,
which take place at lower and higher redshifts (or lower and higher luminosities),
respectively. \citet{Mullaney2012} caution about the effects of both the X-ray (i.e.
AGN) and the infrared (i.e. star formation) luminosities increasing with redshift,
which could mimic a correlation between those values, especially in samples
spanning orders of magnitudes in both $L_{\rm x}$ and $L_{\rm IR}$, and find no clear
signs of a correlation between $L_{\rm x}$ and $L_{\rm IR}$ in moderate luminosity
AGNs ($L_{\rm x}=10^{42}-10^{44}\,{\rm erg\,s^{-1}}$). More recently,
\citet{Mullaney2012b} do find hints of coeval growth of the super-massive black
hole and the host galaxy suggesting a causal connection
\citep[see also][]{Rosario2012}.

In this paper we use the deepest observations from {\it XMM-Newton} and
{\it Herschel}, combined with {\it Chandra} positions and deep multi-wavelength
data in the CDFS to investigate the AGN-host connection, expanding to the less
well-sampled region of high X-ray luminosities
($L_{\rm  x}>10^{44}\,{\rm erg\,s^{-1}}$) and redshifts ($z>2.5$). We exploit the
multi-wavelength information implementing an accurate SED decomposition 
technique to disentangle the AGN and star-formation signals in the optical and
infrared bands, and therefore obtain unbiased star-formation rates for the AGN
sample. We also make use of accurate {\it XMM-Newton} spectra from the deepest
3\,Ms observation for the first time, to investigate the nature of the
AGN - star-formation relation.

\section{Data}

\subsection{X-rays}

Our X-ray data come from the 3\,Ms CDFS {\it XMM-Newton} survey. Initial results of
the survey are presented in \citet{Comastri2011}, and details on the data analysis
and source detection will be presented in Ranalli et al. (in preparation). Briefly, the
bulk of the X-ray observations were made between July 2008 and March 2010, and
have been combined with archival data taken between July 2001 and January 2002,
using a single pointing, and covering a total area of $30\times35$\,arcmin, centred
at the \emph{Chandra} pointing of the CDFS. The total integration time of useful data
is $\rm \simeq2.82\,Ms$. Standard  {\it XMM-Newton} software and procedures were
implemented for the analysis of the data, yielding a point-spread function (PSF)
FWHM of $\approx10.5$\,arcsec, which does not show a significant variation with
the off-axis angle. The XMM-CDFS main catalogue contains 337 sources detected in
the 2-10 keV band with a $>4\sigma$ significance, plus a list of 74 supplementary
sources (detected with PWXDetect, but not with EMLDetect), down to a flux limit of
$\sim 6.6\times 10^{-16}$ erg s$^{-1}$ cm$^{-2}$. X-ray spectra are produced for 169
sources from both lists, detected with a significance above $8\sigma$ and a flux
limit of $\sim 2\times 10^{-15}$ erg s$^{-1}$ cm$^{-2}$. The spectra have been fitted
in {\sc XSPEC} with a simple baseline model of an absorbed power-law and the
addition, if necessary, of a soft excess component and an Fe\,K$\alpha$ line.

\subsection{Optical - near-Infrared}

The area around the CDFS is one of the best observed areas in the sky, with a wealth
of data. In this work, for the identification of our sources in the optical and near-IR
wavelengths we use the MUSYC catalogues of \citet{Gawiser2006} and
\citet{Taylor2009}. \citet{Gawiser2006} present the optical survey of the extended
CDFS \citep[see][hereafter E-CDFS]{Lehmer2005} with the MOSAIC\,II camera of the
4--m CTIO telescope, using a $BVRIz'$ filter set. The source extraction is done using
a combined $BVR$ image and the catalogue is complete to $R_{\rm AB}=25$.
\citet{Taylor2009} combine a large set of optical data, including the
\citet{Gawiser2006} data-set, with near-IR data, primarily from the ISPI instrument
on the CTIO telescope. The catalogue contains sources detected in the $K$ band
down to a $5\,\sigma$ limit of $K_{\rm AB}=22$ and includes photometry in the
$UU_{38}BVRIz'JHK$ bands.

\subsection{Mid-Infrared}
\label{mid-ir}

The entire MUSYC area has been imaged with {\it Spitzer}-IRAC in four bands, 3.6,
4.5, 5.8, and $\rm 8.0\,\mu m$. The central region is imaged as part
of the GOODS survey, and these data are combined with more recent observations of
the wider E-CDFS area in the SIMPLE survey \citep{Damen2011}. The combined
data-set has a $5\,\sigma$ magnitude limit of $\rm [3.6\,\mu m]_{AB}=23.86$,
while the $3\,\sigma$ magnitude limit of the central GOODS region is
$\rm [3.6\,\mu m]_{AB}=26.15$.

The GOODS area in the centre of the CDFS has been imaged with {\it Spitzer}-MIPS in
the $\rm 24\,\mu m$ band with a $5\,\sigma$ flux density limit of
$\rm 30\,\mu Jy$. A much wider area, including the entire E-CDFS was imaged as
part of the FIDEL legacy program \citep[PI: Dickinson; description in][]{Magnelli2009}
with a $5\,\sigma$ flux density limit of $\rm 70\,\mu Jy$; we use a combination of
the two data-sets for this work.

\subsection{Far-Infrared - sub-mm}

The entire E-CDFS region has been imaged with \emph{Spitzer} MIPS in the
$\rm 70\,\mu m$ band as part of the FIDEL survey. For the inner part of the field we
also use the combination of observations from the GOODS-{\it Herschel} survey
\citep{Elbaz2011} and the PACS Evolutionary Probes programme \citep{Lutz2011}.
This combination provides the deepest survey of \emph{Herschel} using the PACS
instrument (Poglitsch et al., 2010) in both the 100 and the $\rm 160\,\mu m$ bands,
with a total integration time of more than 400 hours. Because
GOODS-\emph{Herschel} observations cover only a $13'\times11'$  field inside the
GOODS area, the combined GOODSH-PEP observation has inhomogeneous coverage
of the GOODS-S field, with the GOODS-S outskirt being 2 times shallower than the
inner deep area. The data reduction and image construction procedures for the FIDEL
and the GOODS-{\it Herschel} surveys are described in detail in \citet{Magnelli2009} and
Magnelli et al. (in prep; but see also \citealt{Elbaz2011} and \citealt{Lutz2011}),
respectively. For the source identification and flux density determination, all images
(MIPS-70 and PACS) were treated in a consistent way: MIPS and PACS flux densities
were derived with a PSF fitting analysis, guided using the position of sources
detected in the deep MIPS-24 observations described in Sect.\ \ref{mid-ir}. This
method, presented in detail in \citet{Magnelli2009, Magnelli2011}, has the advantage
that it deals with a large part of the blending issues encountered in dense fields and
provides a straightforward association between MIPS and PACS sources. This
MIPS-24-guided extraction is also very reliable for the purpose of this study,
because in the GOODS-S field the MIPS-24 observations are deep enough to contain
all the AGNs of the MIPS-70 and PACS images \citep{Magnelli2011,Magdis2011}. The
flux density limits of the MIPS-70 catalogue used is 2.5\,mJy ($6\,\sigma$). For the
PACS 100 and $\rm 160\,\mu m$, flux density limits of our catalogues is 0.6 and
1.2\,mJy ($3\,\sigma$) in the $13'\times11'$ inner part of the GOODS-S field and
1.2 and 2.4\,mJy ($3\,\sigma$) in the outskirt of the field. All these values include
confusion noise. For the sub-mm part of the spectrum, we also use the 870\,m
LABOCA and 1.1\,mm AzTEC catalogues of \citet{Weiss2009} and \citet{Scott2010},
which reach depths of 3.5\,mJy and 1.4\,mJy at the 3.7 and $3.5\,\sigma$ levels,
respectively.

\subsection{Radio}

The E-CDFS has been observed with the VLA in two bands (20 and 6\,cm) and the
catalogues are presented in \citet{Kellermann2008} and \citet{Miller2008}, the
former presenting both the 20 and 6\,cm results, and the latter presenting the deeper
20\,cm catalogue. The $5\,\sigma$ flux density limit of the survey near the centre of
the field is $\rm 43\,\mu Jy$ and $\rm 55\,\mu Jy$, at 20\,cm and 6\,cm,
respectively. For this work we also check the much wider and shallower ATCA 20\,cm
observations of \citet{Norris2006}, but we do not find any new identifications of
X-ray sources, however we do find some unique spectroscopic redshift
measurements from their follow-up program; see Sect.\ \ref{specz}. We
also use the VLBI catalogue of \citet{Middelberg2011} to identify any high
surface-brightness VLBI cores among the radio detections, suggestive
of high surface-brightness AGN cores.

\subsection{Redshifts}
\label{specz}

There are a number of spectroscopic campaigns of the CDFS and the E-CDFS
available in the literature. For the purposes of this paper we use spectroscopic
redshifts from the following works: \citet{Balestra2010}; \citet{Casey2011};
\citet{Cooper2011}; \citet{Kriek2008}; \citet{LeFevre2004}; \citet{LeFevre2005};
\citet{Mignoli2005}; \citet{Norris2006}; \citet{Ravikumar2007};
\citet{Silverman2010}; \citet{Szokoly2004}; \citet{Taylor2009}; \citet{Treister2009};
\citet{vanderWel2005} and \citet{Vanzella2008}. For sources which have no
spectroscopic redshift determination we use photometric redshift estimates from
\citet{Cardamone2010a} who use up to 32 optical and infrared bands, including 18
medium narrow-band filters, for $BVR$-detected sources in the E-CDFS. In cases
where the redshift is not available in the \citet{Cardamone2010a} catalogue, or it is
flagged as low-quality, we use the photometric redshifts of \citet{Taylor2009} using
10 bands on $K$-selected sources, \citet{Rafferty2011} who use publically available
photometric catalogues to determine the photometric redshifts of E-CDFS sources,
and \citet{Luo2010} who use up to 35 bands from public catalogues to derive
redshifts of counterparts of {\it Chandra} 2--Ms sources. The typical scatter of the
photometric redshifts is $\Delta z/(1+z)\approx0.01$, and using this value, we
estimate that the induced uncertainty in the infrared luminosities and stellar masses
(see Sects.\ \ref{stellar_masses} and \ref{infrared_luminosity}) from the photometric
redshift uncertainty is $<20\%$, therefore not important for the overall uncertaintie
of the aforementioned values.
\\

\begin{figure}
  \resizebox{\hsize}{!}{\includegraphics{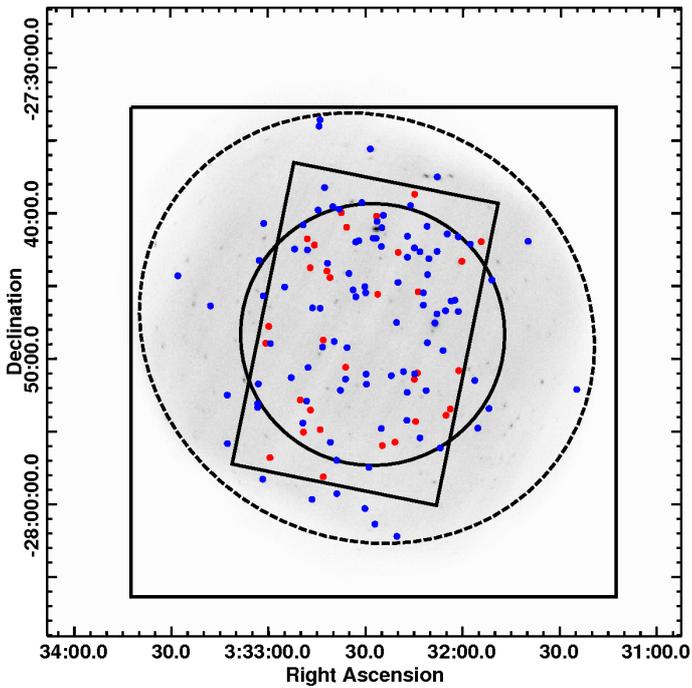}}
  \caption{Spatial limits of the different surveys used in this work. The grey-scale
                 image is the combined 2--10\,keV 3\,Ms image of the {\it XMM-Newton}
                 observations, and the regions are the {\it Herschel} area (small rectangle),
                 the 4\,Ms {\it Chandra} area (solid circle), the {\it XMM-Newton} area used
                 (dashed circle) and the E-CDFS {\it Chandra} area (large square). The radio
                 and {\it Spitzer} areas used are all wider than the E-CDFS. The
                 \emph{Herschel} and the \emph{XMM-Newton} areas are the boundaries of
                 the ``complete'' and ``broad'' samples, respectively (see
                 Sect.\ \ref{final_sample}). The sources of the ``broad'' sample are marked
                 with blue symbols, whereas the sources with FIR flux density upper limits
                 in the ``complete'' sample are marked with red symbols.}
  \label{regions}
\end{figure}

The spatial limits of the different surveys described in this section are shown in
Fig.\ \ref{regions}. The grey-scale image is the combined 2--10\,keV image of
{\it XMM-Newton} and the regions are the \emph{Herschel} area limiting the
combined GOODS-\emph{Herschel}--PEP catalogue (small rectangle), the 4\,Ms
{\it Chandra} area (the region where the effective exposure is larger than half of its
maximum value; solid circle), the {\it XMM-Newton} area (the region where the total
integration time is higher than 1\,Ms; dashed circle) and the E-CDFS {\it Chandra}
area (large solid square). The radio and {\it Spitzer} areas described above are all
wider than the E-CDFS. In this study, we use X-ray sources spanning the entire
\emph{XMM-Newton} region, and use information from all the other surveys to
measure their star-formation rates and stellar masses. The smaller \emph{Herschel}
area is used to construct a ``complete'' sample of X-ray AGNs, where we have FIR
detections or upper limits for the majority of the AGNs (see Sect.\ \ref{final_sample}),
while for the wider area (``broad'' sample) we use $\rm 70\,\mu m$ measurements
from the FIDEL survey.

\section{The sample}

In order to avoid high X-ray flux spurious detections due to relatively high
background levels, we limit the sample to those sources which have a combined
\emph{XMM-Newton} exposure of 1\,Ms or higher in the 2--10\,keV band
(356 sources in the main and supplementary catalogues of Ranalli et al. in
preparation). To better constrain their positions we look for counterparts among the
X-ray sources observed with the {\it Chandra} surveys, namely in the 2\,Ms CDFS
catalogue of \citet{Luo2010}, the 4\,Ms CDFS catalogue of \citet{Xue2011}, and the
E-CDFS catalogues of \citet{Lehmer2005} and \citet{Virani2006}. The characteristic
positional uncertainty of {\it Chandra} is $\lesssim1$\,arcsec, compared to the
4--5\,arcsec of {\it XMM-Newton}. We keep {\it Chandra} counterparts which are
within 5\,arcsec of the {\it XMM-Newton} position and find 311 unique associations.
We also look for counterparts in the $\rm 3.6\,\mu m$ SIMPLE catalogue, using the
likelihood ratio method\footnote{The likelihood ratio method \citep{Sutherland1992}
is usually adopted in cases where a counterpart is sought in a crowded catalogue (in
this case the SIMPLE catalogue), and it uses the surface density of objects of a given
magnitude to estimate the probability that a counterpart at a certain distance is a
chance encounter. An example of using this method to find infrared counterparts of
\emph{XMM-Newton} sources can be found in \citet{Rovilos2011}} with a matching
radius of 5\,arcsec, and find another 19 sources with $LR>0.85$ and all with a
reliability $>99.9\%$\footnote{The reliability is a measure of the probability that the
selected counterpart is the correct one, and it is used in cases where more than one
possible counterparts are found.}. Most of them lie in the area not covered by the
4\,Ms {\it Chandra} CDFS survey (see Fig.\ \ref{regions}). Our final X-ray catalogue
contains 330 sources with good positional constraints ($\lesssim 1$\,arcsec) either
from {\it Chandra}, or from {\it Spitzer}-IRAC. Of the 26 sources with no
unambiguous \emph{Chandra}, or \emph{Spitzer}-IRAC counterpart, 23 are of
low-significance ($<5\,\sigma$) and therefore likely spurious, and three are double
sources in \emph{Chandra}, not resolved by \emph{XMM-Newton}, which we exclude
from our sample.

Next, we build the multi-wavelength catalogue of the X-ray sources using all the
information available and the likelihood ratio method to select the counterparts,
using the positional uncertainties provided in the various catalogues. We first
combine the SIMPLE catalogue with both the $K$-selected \citep{Taylor2009} and the
$BVR$-selected \citep{Gawiser2006} MUSYC catalogues (preferring $K$-selected
sources in cases where they are detected in both catalogues) and find the
optical-infrared counterparts of the X-ray sources, constraining their positions, and
then we look for counterparts in the FIDEL and $\rm 24\,\mu m$-prior {\it Herschel}
catalogues. We find a counterpart in at least one of the optical or infrared catalogues
for 328/330 X-ray sources; one source is too faint to be detected at any other
wavelength than X-rays and the other is close to a bright optical-infrared source and
is missed by the source detection algorithms. The positions of our optical-infrared
counterparts are in good agreement with those of \citet{Xue2011} and
\citet{Luo2010} for the sources in common; more than 90\% of the counterparts are
within 0.7\,arcsec of the optical positions given in those catalogues. We use the
optical positions to look for radio counterparts in the \citet{Kellermann2008} and
\citet{Miller2008}, and find 53 matches within a 2\,arcsec radius, excluding X-ray
sources which have multiple radio counterparts within 10\,arcsec, which would be
indicative of FR\,II radio AGNs. For the sub-mm catalogues, because of their large
positional uncertainties ($\sim8$\,arcsec) we use the likelihood ratio method to
assign a FIDEL-24 counterpart to the sub-mm sources, and if this is the same as the
FIDEL-24 counterpart of the X-ray source we consider it as reliable. We find a
LABOCA counterpart for five X-ray sources, and an AzTEC for two of these five.
Finally, we look for redshifts and find 215 spectroscopic redshift determinations
from the various catalogues listed in \S\,\ref{specz}, and 106 photometric redshift
estimates. Nine sources have no redshift determination, because they are too faint at
optical wavelengths. In the final catalogue we also include FIR upper limits for X-ray
sources which are in the area observed by PACS (see Fig.\ \ref{regions}) with no
detection. There are 155 {\it XMM-Newton} sources inside the PACS area and 94 of
them are detected. Of the remaining 61, 20 are in regions confused with nearby
bright FIR sources, and 41 are upper limits. 

\subsection{Stellar masses}
\label{stellar_masses}

The most reliable method to derive stellar masses for galaxies is the fitting of their
broad-band spectral energy distributions (SEDs) with synthetic stellar templates with
known star-formation histories and dust extinction properties
\citep[see][for the limitations of the method]{Shapley2001,Papovich2001}. The stellar
component is important at optical wavelengths $\rm (\lesssim1\,\mu m)$, but we
use the full multi-wavelength information (excluding radio and X-rays) to fit the
SEDs. The reason for this is that the AGN can affect the optical properties of the
system \citep[see e.g.][]{Pierce2010}, and by fitting a combination of AGN and host
templates using the infrared photometric information we can constrain the AGN
contribution.

\begin{figure}
\subfloat{\label{328}\resizebox{43mm}{!}{\includegraphics{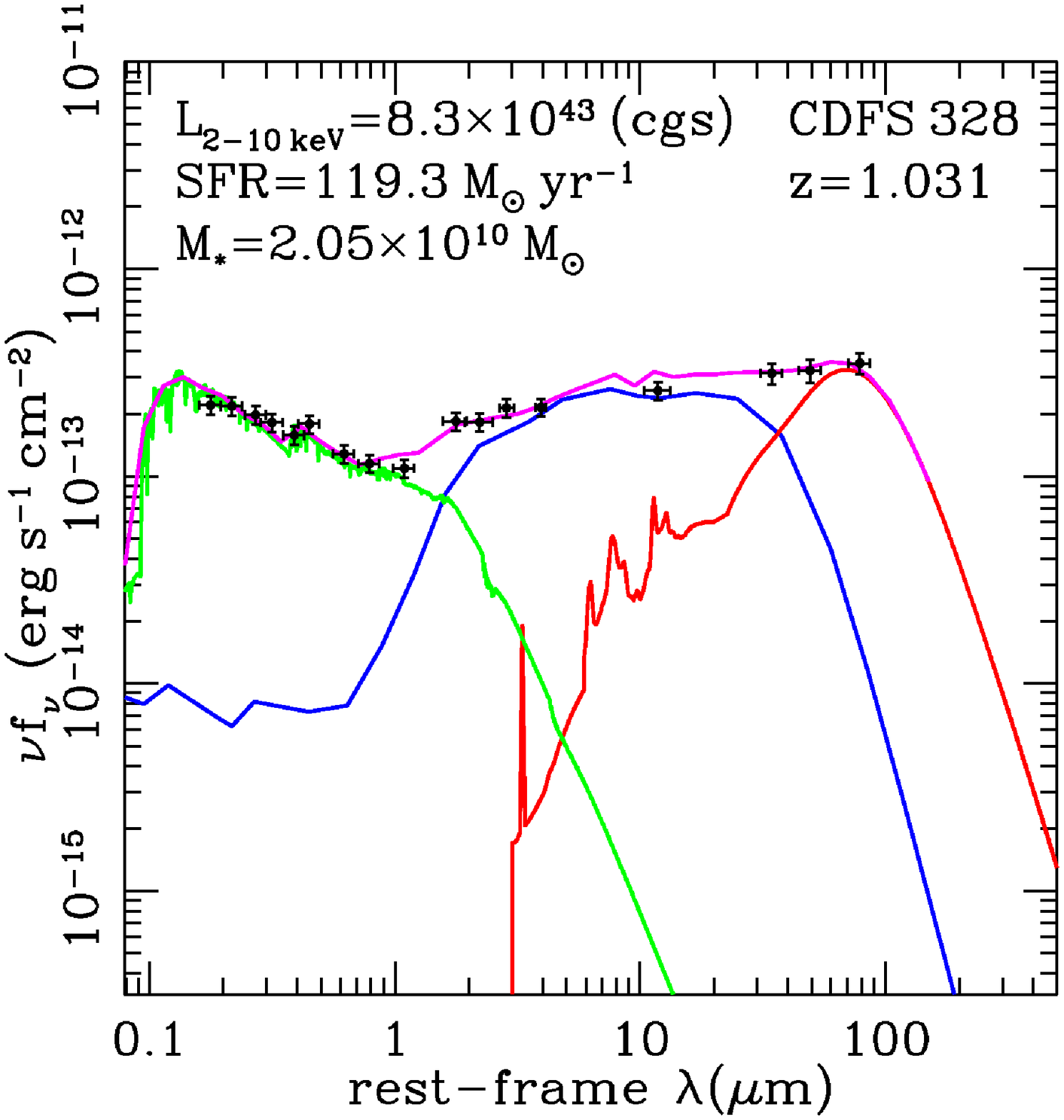}}}
\hspace{2mm}
\subfloat{\label{243}\resizebox{43mm}{!}{\includegraphics{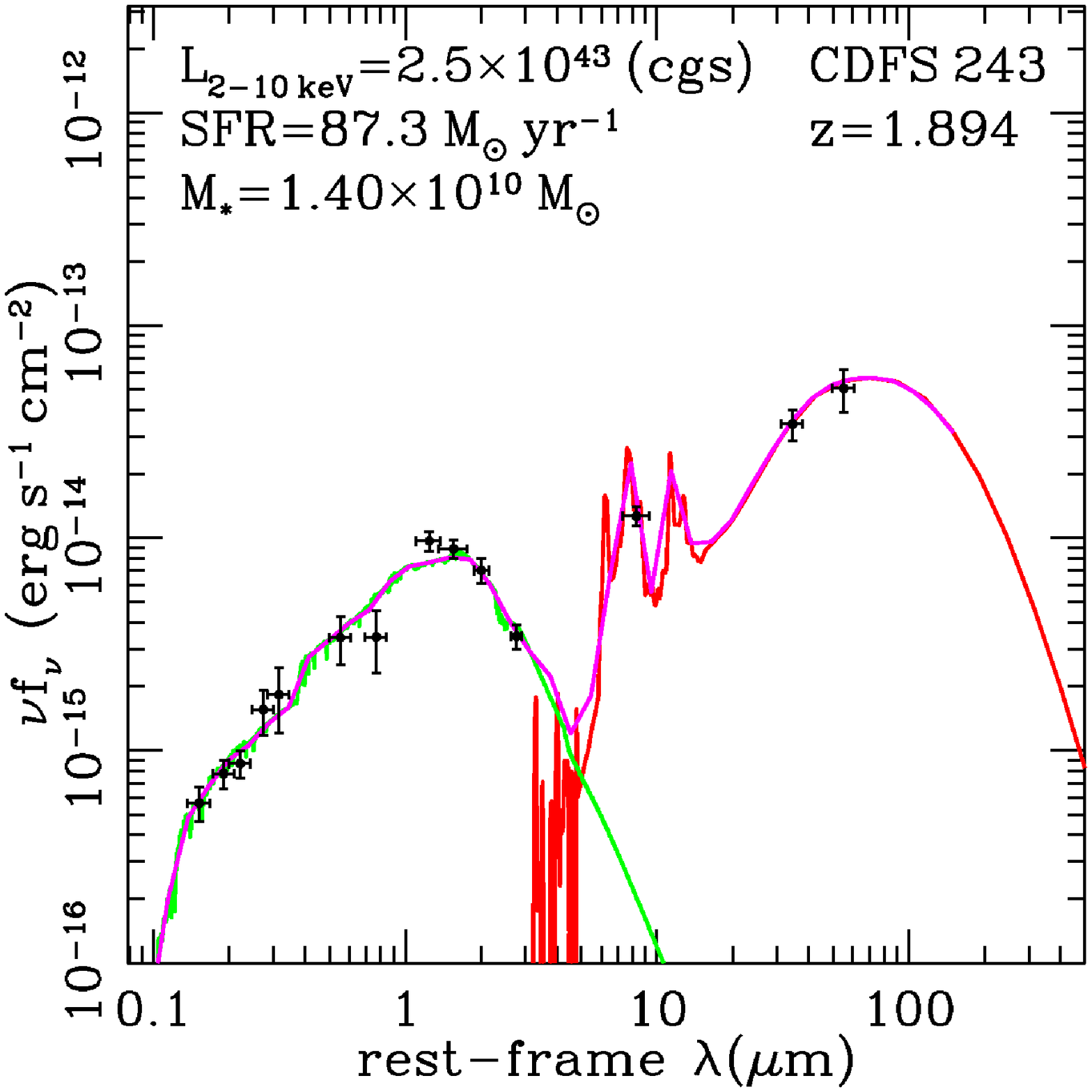}}}
\\
\subfloat{\label{228}\resizebox{43mm}{!}{\includegraphics{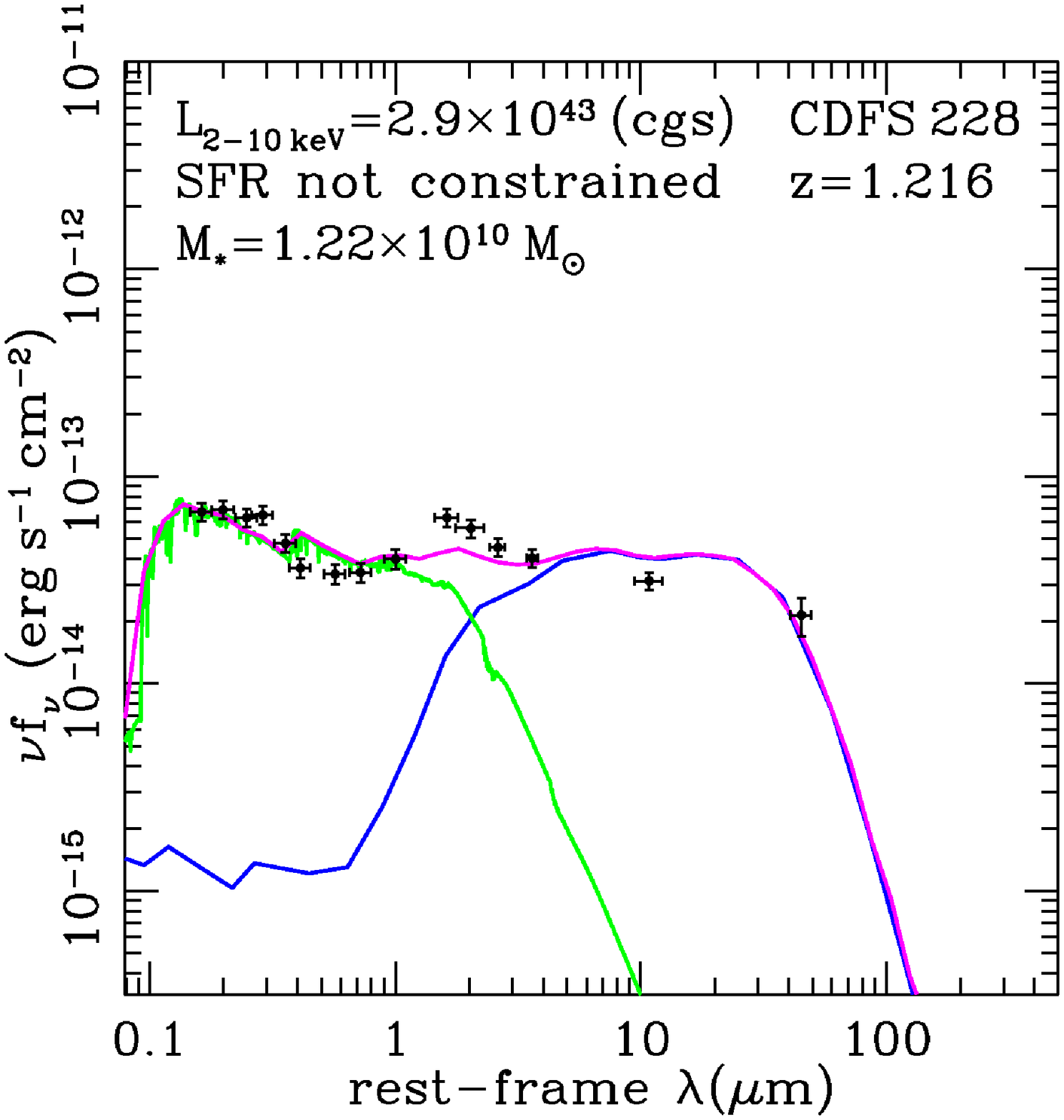}}}
\hspace{2mm}
\subfloat{\label{153}\resizebox{43mm}{!}{\includegraphics{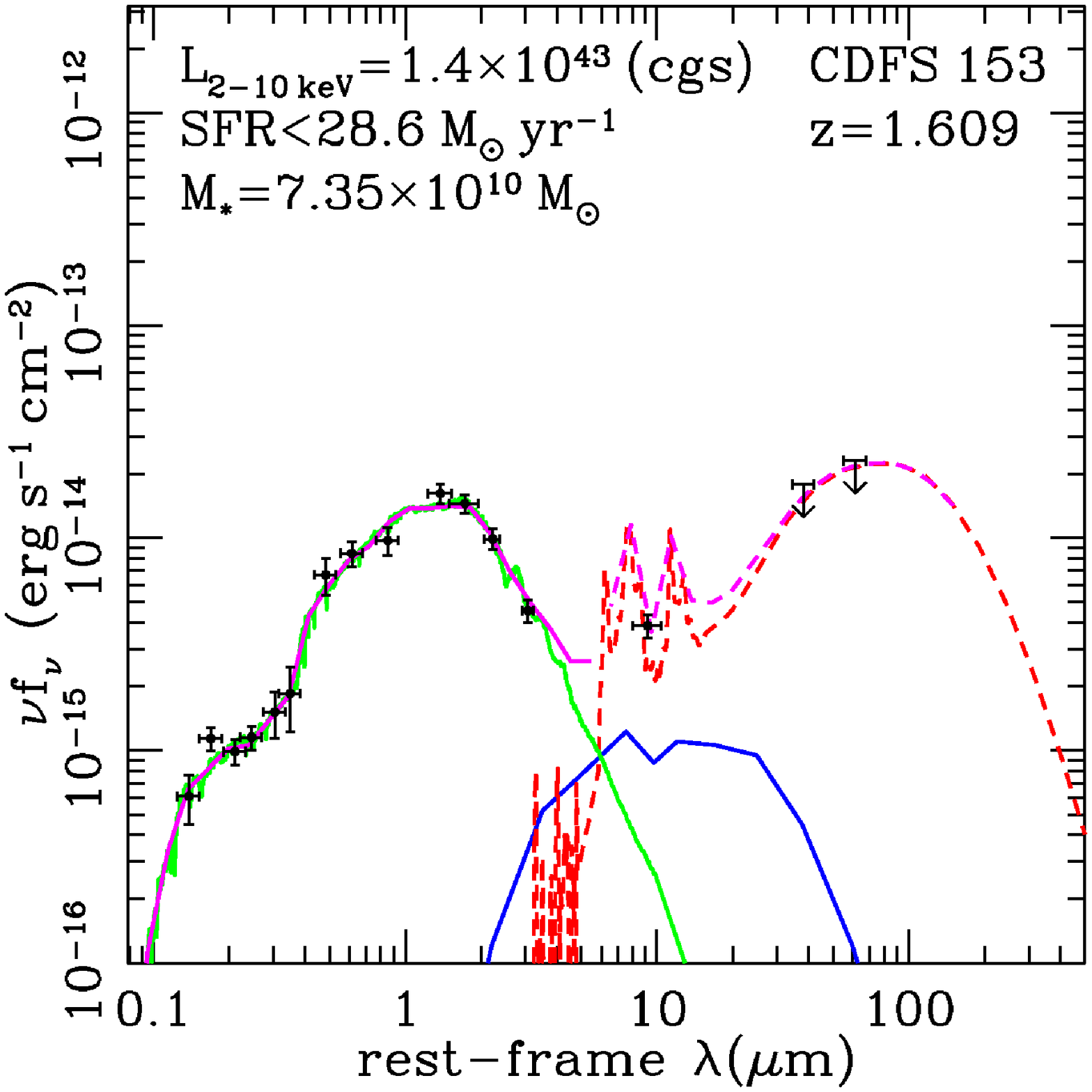}}}
  \caption{Examples of the SED fitting used to derive stellar masses and
                star-formation rates, selected to demonstrate the diversity in the SEDs of
                the X-ray sample. The star-formation component is plotted in red, the
                AGN component in blue, and the stellar component in green. In magenta
                we plot the combination fitted to the data-points. The stellar component is
                only confined by the optical wavelengths where it usually dominates the
                flux, but there are cases where we detect a substantial AGN contribution.
                The star-formation component is confined by the FIR flux with rest-frame
                wavelength $\rm >20\,\mu m$, and again there are cases where the FIR
                flux is dominated by the AGN. With downward arrows we plot far-IR upper
                limits (see Sect.\ \ref{final_sample}), where the resulting starburst
                component is plotted with a dashed line. The numbering in the top-right
                corners of the panels refers to the preliminary {\it XMM-Newton} catalogue
                number.}
  \label{opt_SEDs}
\end{figure}

For the optical-infrared SED fitting we use the procedure described in
\citet{Lusso2011}. We apply a $\chi^{2}$ minimisation method using stellar
templates from the \citet{Bruzual2003} stellar synthesis code, applying solar
metallicity, a mixture of constant and exponentially decaying star-formation rates,
and a Galactic disk initial mass function \citep[IMF;][]{Chabrier2003}. We redden the
stellar SEDs using the \citet{Calzetti2000} law, and combine the reddened SEDs with
star-formation infrared SEDs from \citet{Chary2001} (105 templates with different
FIR profiles in the $\rm 3-1000\,\mu m$ range) and four AGN SEDs from
\citet{Silva2004}, which span from the optical to the far-infrared with different
absorption properties (unabsorbed to $N_{\rm H}\gtrsim10^{24}\,{\rm cm^{-2}}$). Some
characteristic results of the  optical-to-infrared SED fitting, as well as a FIR-limit
example can be seen in Fig.\ \ref{opt_SEDs}. These are examples of both AGN and
starburst dominated SEDs. There is enough optical information (photometry and
redshift) to fit an SED for 304 of the 330 sources. Sources with a detection in the
\citet{Taylor2009} catalogue (the majority of the optically-detected sources) typically
have photometry in nine optical - near-IR bands\footnote{We do not use the
photometry in the $U_{38}$ band for the SED fitting}, and those detected only in the
\citet{Gawiser2006} in six bands, which are used for the determination of the stellar
masses. We do not take into account upper limits for the fitting, but check that the
predicted flux  density value of the fitted SED is indeed lower than the limit. The
reduced $\chi^2$ values of the best-fit models are typically in the 1--10 range, after
reprocessing the flux density errors using a quadratic combination with the
10\%-level error, to account for the typical flux differences between the SED
templates used.

The method we use to calculate the stellar masses induces uncertainties both from
the choice of the different parameters fitted, and from the $\chi^2$ procedure itself.
The derived stellar mass values are potentially strongly influenced by such
uncertainties, especially at high redshifts, like the majority of the sources in our
sample \citep[see also][]{Michalowski2012}. The uncertainties from the parameters
included in the stellar synthesis procedure are estimated to be $\sim0.15$\,dex
\citep{Bolzonella2010}, and checking the stellar masses and $\chi^2$ values of SED
fits with different templates and different relative contributions, we estimate the final
uncertainty in the stellar masses to be $\sim0.25$\,dex at the 90\% confidence level
\citep[see] [for a more detailed description of the method]{Lusso2012}. We also note
that the use of the \citet{Chabrier2003} IMF causes a slight underestimation of the
stellar masses with respect to the \citet{Kroupa2001} IMF, in particular they are on
average lower by a factor of $\sim1.1$
\citep[see][]{Bolzonella2010,Pozzetti2010,Hainline2011}. In this work we use the
\citet{Chabrier2003} IMF in order to avoid an over-prediction in the number of
low-mass stars \citep{Hainline2011}, but we also combine the stellar masses with
star-formation rates; the latter are based on infrared luminosities. This
star-formation rate proxy uses the \citet{Kroupa2001} IMF for its calibration
\citep[see][]{Murphy2011}, so we increase the stellar masses we derive through the
SED fitting by a factor of 1.1 to be consistent with the star-formation rates.

\subsection{Star-formation rates}

Star formation in galaxies affects almost all of their observed properties, from the
X-rays to the radio wavelengths, so there are traditionally a number of ways to
measure the star-formation rate (SFR). In the cases of AGN hosts we can rule out the
X-rays, since they are completely outshone by the AGN (see also
\S\,\ref{final_sample}). In this work we test three methods based only on flux density
measurements: i) infrared luminosity, ii) radio luminosity, and iii) optical SED fitting.

\subsubsection{Infrared luminosity}
\label{infrared_luminosity}

The IR luminosity is arguably the most reliable tracer of star-forming activity and is
well correlated with other tracers
\citep[see][for reviews]{Kennicutt1998a,Kennicutt2012}. The IR photons are emitted
by the dust surrounding young stars, which is heated by their ultra-violet radiation.
In this paper we will use the integrated rest-frame $\rm 8-1000\,\mu m$
luminosity and the equation:
\begin{equation}
\left(\frac{\rm SFR}{\rm M_{\sun}\,yr^{-1}}\right)=3.88\times10^{-44}\left(\frac{L_{\rm IR}}{\rm erg\,s^{-1}}\right)
\end{equation}
from \citet{Murphy2011}. In order to measure the IR luminosity we perform the SED
decomposition described in \S\,\ref{stellar_masses} anew, using only the infrared
data-points from {\it Spitzer} and {\it Herschel}, the complete \citet{Chary2001} host
templates (i.e. including wavelengths $\rm <3\,\mu m$), the AGN templates, and
ignoring the synthetic stellar part. The method we use is the same as described in
\citet{Georgantopoulos2011a} and \citet{Georgantopoulos2011b}, using the SED
templates described in \S\,\ref{stellar_masses}. We do this in order to avoid the
degeneracies in the optical wavelengths between the stellar and AGN light, which
could affect the fitted AGN contribution to the infrared luminosity; this way we fit
fewer free parameters. For the infrared SED decompositions we require at least three
mid-IR points from \emph{Spitzer}-IRAC to determine the shape of the mid-IR part
of the SED, and at least one flux density determination in a rest-frame wavelength
higher than $\rm 20\,\mu m$, which we can use to constrain the far-IR part of the
SED; there are 125 X-ray sources which comply with these criteria. The FIR flux
density determination comes from FIDEL-$\rm 70\,\mu m$ (for $z<2.5$),
PACS-$\rm 100\,\mu m$ (for $z<4$), PACS-$\rm 160\,\mu m$, sub-mm (LABOCA
and/or AzTEC), or a combination of them. We use the galaxy component (AGN-free)
to calculate the star-formation rates, and for 14 cases the AGN component
dominates even the longest wavelength IR data-point available, so the determination
of the AGN-free part of the IR emission is not reliable. The uncertainties in the IR
luminosity values come mostly from the SED decomposition, and a check of the
$\chi^2$ values of fitting secondary solutions not selected yields an uncertainty of
$\sim0.3$\,dex (or a factor of two) in the 90\% confidence level, for the vast majority
of the sources. The uncertainties arising from the far-IR flux errors are much lower.

\subsubsection{Optical SED fitting}

The star-formation rate can be derived as a by-product of the optical SED fitting
performed in \S\,\ref{stellar_masses}, using the star-formation history and age
assumed, and the normalisation from the photometry. Similar methods have been
widely used in deep fields, including the CDFS \citep[e.g.][]{Brusa2009}, especially if
far-infrared photometry is not available. In the next paragraph we will test its
reliability, since it is a highly model-dependent method with systematic
uncertainties arising mainly from the IMF, star-formation history and extinction law
used \citep[see][]{Bolzonella2010}. 

\begin{figure*}
  \subfloat[FIR - optical SED]{\label{SFR_FIR_opt}\resizebox{90mm}{!}{\includegraphics{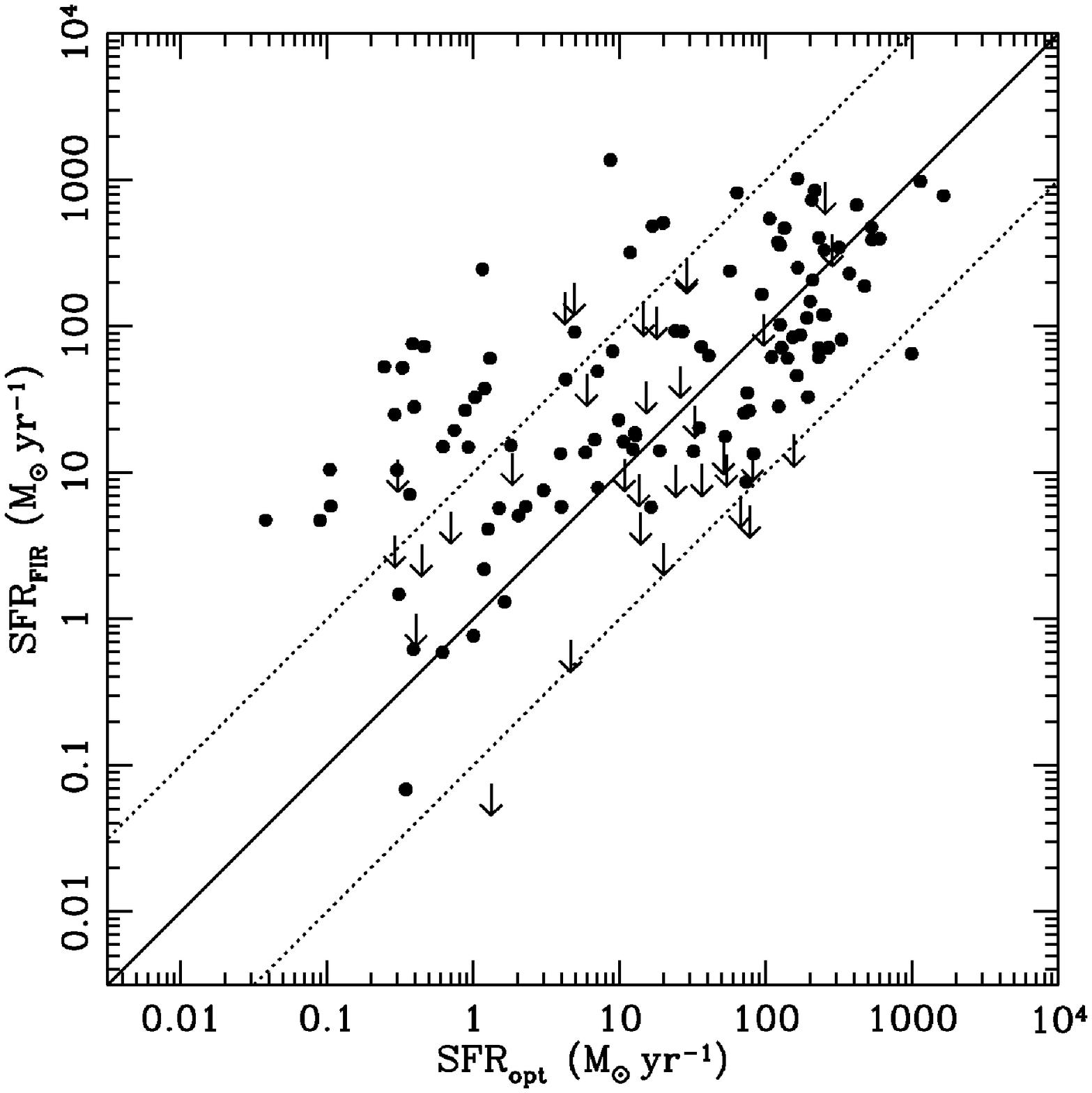}}}
\hspace{4mm}
  \subfloat[FIR - radio]{\label{SFR_FIR_rad}\resizebox{90mm}{!}{\includegraphics{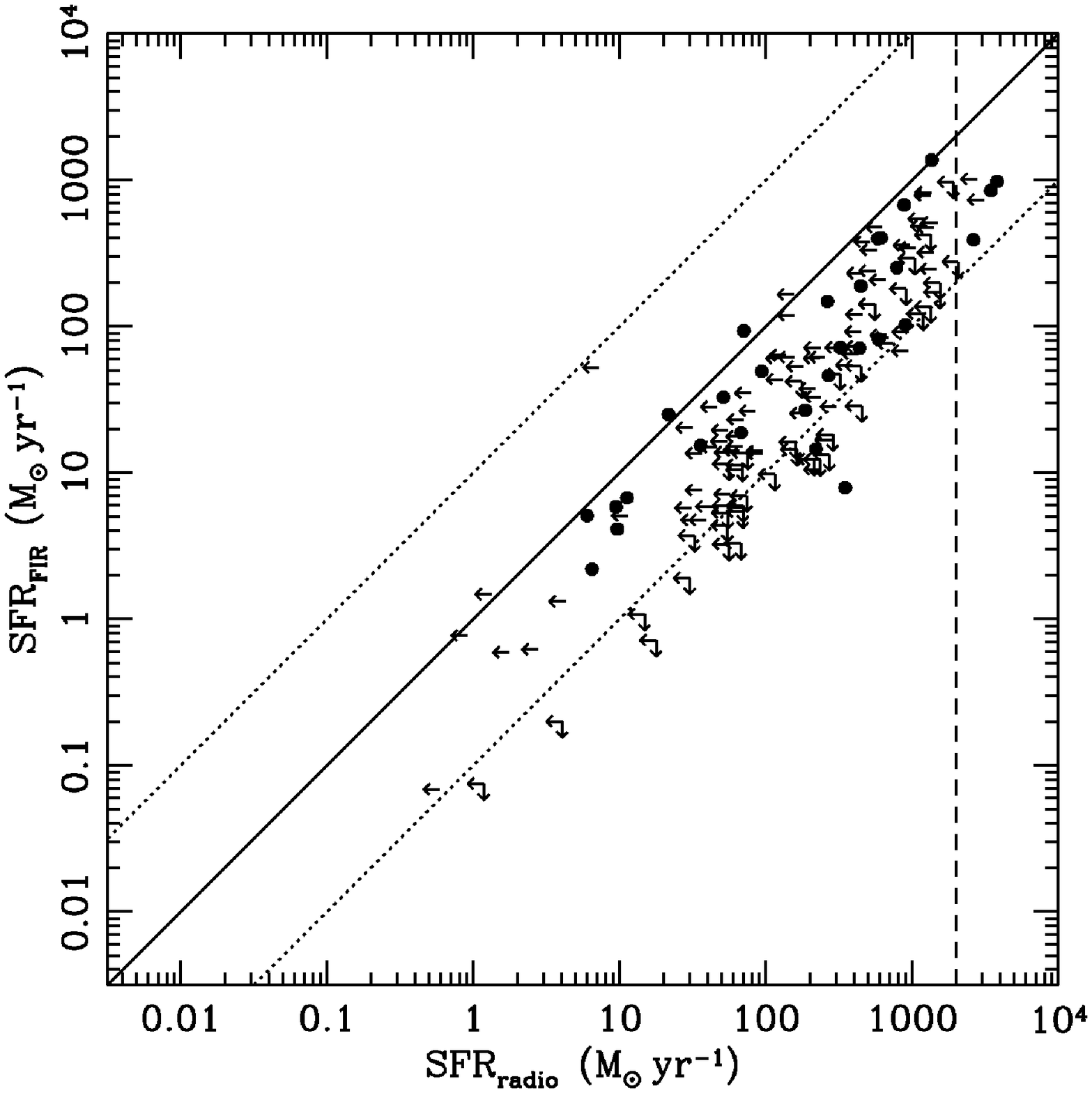}}}
  \caption{Comparison of the star formation rates of X-ray sources, measured using
                 the infrared $\rm 8-1000\,\mu m$ luminosity, and the optical SED fitting
                 (a), or the 1.4\,GHz radio luminosity (b). Black circles mark reliable fits in
                 both optical and far-infrared (or radio) wavelengths, excluding cases
                 where the AGN SED dominates even the longest wavelength flux density
                 available. Sources spanning over  the whole 3\,Ms \emph{XMM-Newton}
                 region are used in these plots, while the downward arrows refer to
                 \emph{Herschel} non-detections in the PACS area (35 cases), and leftward
                 arrows in the right panel refer to radio upper limits spanning over the
                 whole \emph{XMM-Newton} region (112 cases). The solid and dashed
                 lines in both panels represent the 1:1 relation and the $\rm\pm1\,dex$
                 deviation, respectively, and the vertical dashed line in the right panel
                 marks the calculated star-formation rate of a source having
                 $L_{\rm 1.4\,GHz}=10^{24.5}\,{\rm W\,Hz^{-1}}=10^{31.5}\,{\rm erg\,s^{-1}\,Hz^{-1}}$.
                 The significant deviation form the 1:1 relation of both cases and the
                 significant scatter of the optical SED (left) case prevent us from using any
                 of the two SFR tracers in this paper.}
  \label{SFR_compare}
\end{figure*}

In Fig.\ \ref{SFR_FIR_opt} we plot the star-formation rate measured from the infrared
luminosity of the sources detected in the far-infrared against the SFR measured from
the optical SED fitting after correcting for extinction, excluding the AGN contribution
for both cases. All X-ray sources with an infrared measurement with rest-frame
wavelength above $\rm 20\,\mu m$ and an optical identification (in MUSYC) are
plotted with a circle. We exclude these 14 cases where, according to the SED
decomposition, the flux density of the longest wavelength data-point is dominated
by the AGN, so that the SFR cannot be constrained (see Fig.\ \ref{228}). In
Fig.\ \ref{SFR_FIR_opt} we also include SFR upper limits for the X-ray sources in the
\emph{Herschel} area which are not detected by \emph{Herschel}; for one of the 41
X-ray sources with FIR upper limits we do not have any photometric data-points to
perform an SED fitting in the mid-infrared, while another four are not detected in the
optical, or have no redshift determination, so 36 upper limits are plotted in
Fig.\ \ref{SFR_FIR_opt}. The solid line is the 1:1 line and the dotted lines mark the
$\pm1$\,dex region. We can see that out of the 109 points of Fig.\ \ref{SFR_FIR_opt},
79 are between the dotted lines, while for 29 sources using the optical SEDs
underestimates the SFR by more than an order of magnitude; for one (and two upper
limits) the SFR is overestimated by more than an order of magnitude, assuming that
the infrared SFR is reliable. The star-formation rate estimated from the optical SED is
a highly model-dependent value, and is very sensitive to the star-formation history
assumed in the stellar synthesis models, which shape the optical SED. It is also
sensitive to dust extinction, which would cause an underestimation of the SFR,
explaining the behaviour we see in Fig.\ \ref{SFR_FIR_opt}. Due to this large scatter
and systematic offset, we do not rely on the optical SED fitting to derive
star-formation rates and use it only for stellar mass determinations. The stellar mass
is an integrated value and therefore less sensitive to the assumed star-formation
history. Moreover, we do not detect any obvious dependence of the difference
between the SFR determination using the two estimators on X-ray or optical classes,
which would indicate AGN contamination as the cause of the scatter.

\subsubsection{Radio luminosity}

The radio luminosity of star-forming galaxies is tightly correlated with their infrared
luminosity \citep[see][for a review]{Condon1992}, and this correlation holds even for
cosmologically significant redshifts
\citep[$z\approx2$;][]{Appleton2004,Ivison2010}; we test it in this work as a
possibility to derive the star-formation rates in X-ray sources without far-infrared
detections. The radio emission in star-forming systems is generated by synchrotron
radiation from relativistic electrons accelerated by supernova induced shocks and
free-free emission in \ion{H}{ii} regions. The caveat is that AGNs themselves can
produce radio emission through radio jets or compact high surface-brightness
synchrotron core emission from relativistic electrons heated by the AGN. There is a
dichotomy in the radio power of quasars \citep{Miller1990}, with sources having
$L_{\rm 5\,GHz}\gtrsim10^{25}\,{\rm W\,Hz^{-1}}$ being characterised as ``radio-loud'' 
and their power source being closely connected to the AGN, and sources having
$L_{\rm 5\,GHz}\lesssim10^{25}\,{\rm W\,Hz^{-1}}$ being characterised as
``radio-quiet'' and having a controversy about their power source. Alternatively, the
radio to optical flux ratio is used in some studies to differentiate between
radio-quiet and radio-loud AGNs \citep[][]{Kellermann1989}. However, the dichotomy
between radio-loud and radio-quiet sources is not clear if more complete samples
are used \citep[e.g.][]{White2000}, with many objects in the ``intermediate'' region,
making the transition smooth with only a vague limit. Recently, \citet{Padovani2011},
using the luminosity functions of different types of radio sources in the CDFS, argue
that the major contribution to radio power in the diffuse part of radio-quiet AGNs
comes from star-formation, though taking a somewhat stringent limit to
characterise the radio sources based on their radio luminosities
($L_{\rm 1.4\,GHz}=10^{24.5}\,{\rm W\,Hz^{-1}}=10^{31.5}\,{\rm erg\,s^{-1}\,Hz^{-1}}$,
combined with other observational characteristics).

For this work we test how reliable the radio luminosities are in estimating
star-formation rates for an AGN sample using the VLA 1.4\,GHz flux densities from
\citet{Kellermann2008} and \citet{Miller2008}, also checking the VLBI catalogue of
\citet{Middelberg2011} to exclude any high surface-brightness compact cores,
characteristic of non-thermal nuclear emission, not connected to the star formation
\citep[e.g.][]{Giroletti2009}. We calculate the radio luminosities using
\begin{equation}
L_{\rm 1.4\,GHz}=4\pi d_{\rm l}^{2}S_{\rm 1.4\,GHz}(1+z)^{\alpha-1}
\end{equation}
where $\alpha$ is the radio spectral index, assuming $S_{\nu}\propto\nu^{-\alpha}$,
and it is calculated from the relative radio flux densities at 1.4 and 5\,GHz. In cases
where the 5\,GHz flux density is not available we assume $\alpha=0.8$,
characteristic of synchrotron emission \citep[see][]{Condon1992}. There are 53 X-ray
sources with a radio counterpart within 2\,arcsec and without another radio source
closer than 10\,arcsec, the latter would suggest an FR\,II radio-loud source. Eight of
these sources have a high surface-brightness core detected with VLBI with a flux
density above 0.5\,mJy, and are removed from the test sample, and a further eight
are radio-loud according to the $L_{\rm 1.4\,GHz}>10^{31.5}\,{\rm erg\,s^{-1}\,Hz^{-1}}$
criterion (three are also detected in the far-infrared and are included in
Fig.\ \ref{SFR_FIR_rad}). The star-formation rate is calculated using
\begin{equation}
\left(\frac{\rm SFR}{M_{\sun}\,{\rm yr}^{-1}}\right)=6.35\times10^{-29}\left(\frac{L_{\rm 1.4\,GHz}}{\rm erg\,s^{-1}\,Hz^{-1}}\right)
\end{equation}
\citep{Murphy2011}. In Fig.\ \ref{SFR_FIR_rad} we plot the star-formation rates from
the infrared and radio luminosities for sources being detected in both bands, keeping
the same range and symbols as in Fig.\ \ref{SFR_FIR_opt}. We also plot the radio and
infrared (for the \emph{Herschel} area) upper limits with arrows. The scatter in this
case is significantly lower than in Fig.\ \ref{SFR_FIR_opt}. However, the mean
$\rm\log(SFR_{IR}/SFR_{1.4\,GHz})$ of the radio detections is $-0.48$ (without taking
into account the VLBI sources and the upper limits) with a standard deviation of
$\sigma=0.39$. The dashed vertical line in Fig.\ \ref{SFR_FIR_rad} marks the
calculated SFR of a source having
$L_{\rm 1.4\,GHz}=10^{31.5}\,{\rm erg\,s^{-1}\,Hz^{-1}}$, thus being border-line
radio-loud according to the limit of \citet{Padovani2011}. If we keep this limit and
calculate the mean $\rm\log(SFR_{IR}/SFR_{1.4\,GHz})$ of only the radio-quite objects,
its mean and standard deviation become $-0.45\pm0.40$, so there is still
contamination from the AGN emission; most probably we are detecting in the radio
band those sources which are in the top of the radio flux distribution, something
which is supported by the location of the upper limits. Because of this
contamination, we do not use the radio power as a star-formation proxy in our AGN
sample.

\subsection{Final sample}
\label{final_sample}

We start with a sample of 356 X-ray 2--10\,keV selected sources from the 3\,Ms
{\it XMM-Newton} survey with a total integration time more than 1\,Ms, 330 of which
have good ($\lesssim1$\,arcsec) positional constraints from {\it Chandra} or
{\it Spitzer}. We have enough optical information to fit an SED and calculate the
stellar masses for 304 of these 330 sources. On the other hand, we rely on the
infrared flux density to constrain the SFR of our sample, and there are 111 sources
with an SFR measurement from the FIR flux. For 109 of them we can also calculate
the stellar mass from the optical SED.

\begin{figure}
  \resizebox{\hsize}{!}{\includegraphics{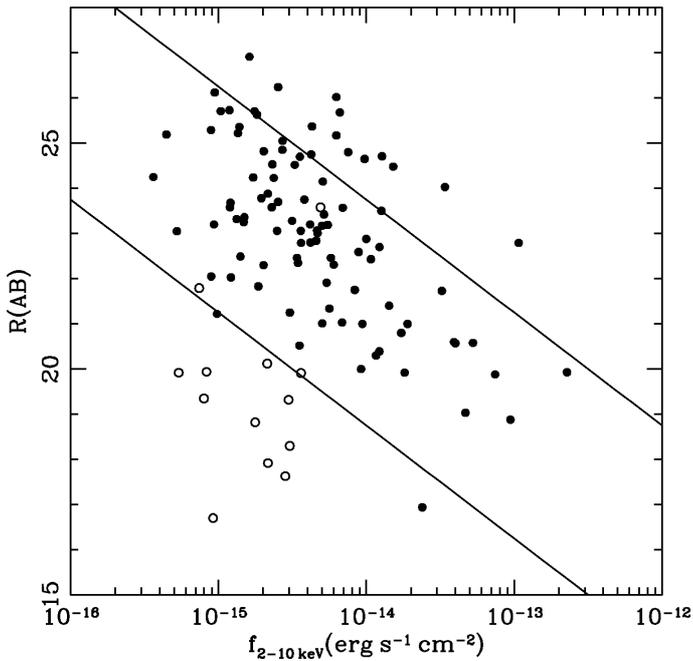}}
  \caption{Optical magnitude against 2--10\,keV flux density for the
                {\it XMM-Newton} sources with robust SFR estimations. The lines mark
                the $-1<\log(f_{\rm x}/f_{\rm opt})<1$ region and open symbols sources with
                $L_{\rm x}<10^{42}\,{\rm erg\,s^{-1}}$. We exclude the 10 sources which
                have both $L_{\rm x}<10^{42}\,{\rm erg\,s^{-1}}$ and
                $\log(f_{\rm x}/f_{\rm opt})<-1$ from our final sample, as normal galaxy
                candidates.}
  \label{fxfo}
\end{figure}

As we are dealing with faint X-ray fluxes, there is the possibility that for some of the
X-ray sources the X-rays trace normal star-forming galaxies instead of the AGNs
\citep{Ranalli2003,Bauer2004}. In Fig.\ \ref{fxfo} we plot the hard X-ray flux
densities against their optical ($R$-band) magnitudes. The lines mark the
$-1<\log(f_{\rm x}/f_{\rm opt})<1$ region where the bulk of the AGNs are expected
\citep[see e.g.][]{Stocke1991,Elvis1994,Xue2011}. Sources with
$\log(f_{\rm x}/f_{\rm opt})<-1$ are candidates for being normal galaxies instead
of AGNs \citep[see][]{Tzanavaris2006,Georgakakis2006}. Moreover, most normal
galaxies have X-ray luminosities not exceeding $\rm 10^{42}\,erg\,s^{-1}$, except for
a few extremely star-forming sources, mainly detected in sub-mm wavelengths
\citep[see e.g.][]{Alexander2005,Laird2010}. Sources with luminosities below the
$\rm 10^{42}\,erg\,s^{-1}$ limit are plotted with open circles in Fig.\ \ref{fxfo}. There
are 10 sources compliant with both the $f_{\rm x}/f_{\rm opt}$ and the $L_{\rm x}$
criteria, and they do not show any signs of obscuration in their X-ray spectra, so we
remove them from the AGN sample. 

\begin{figure}
\subfloat{\label{SFR_hist}\resizebox{43mm}{!}{\includegraphics{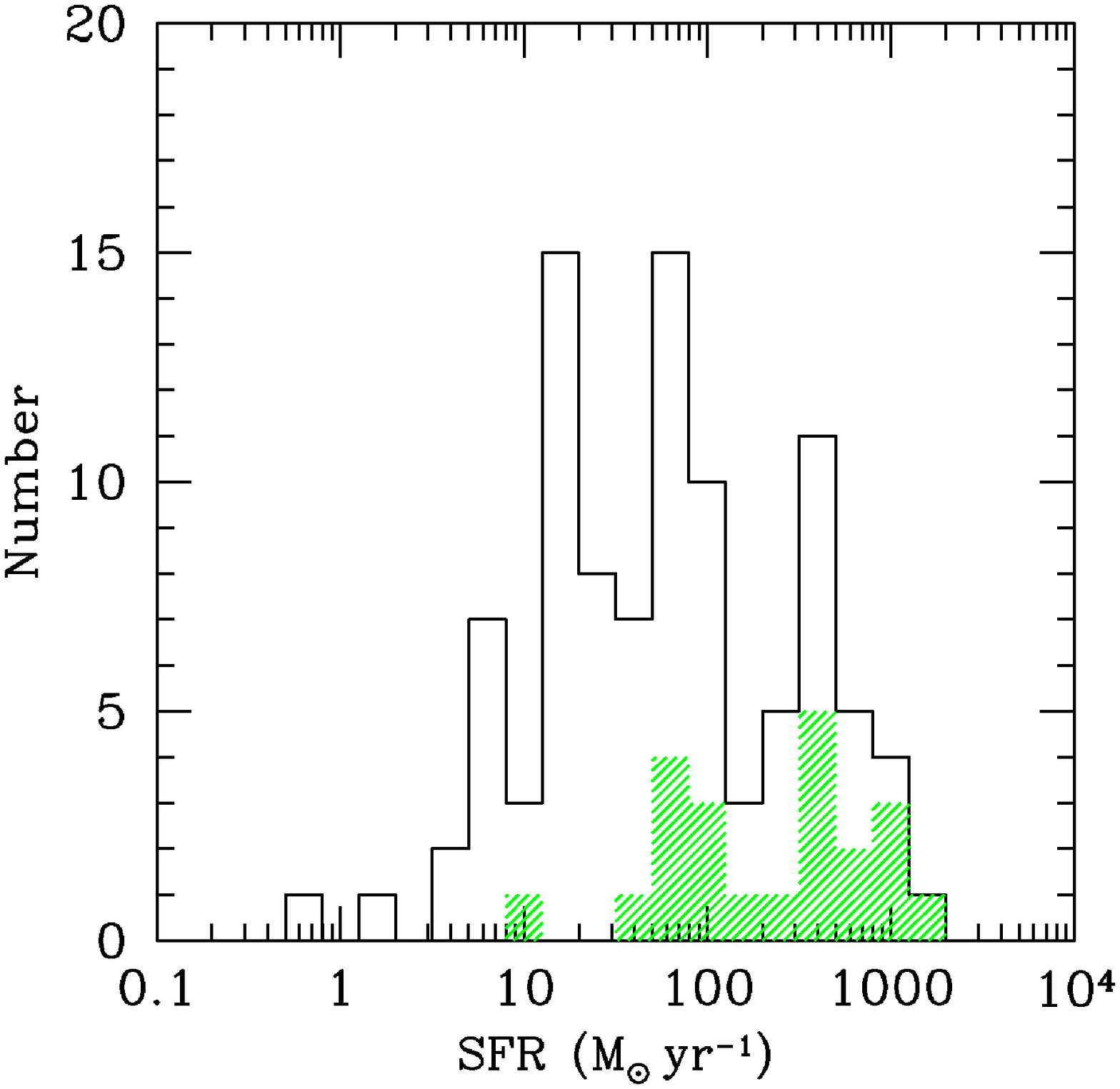}}}
\hspace{2mm}
\subfloat{\label{M_star_hist}\resizebox{43mm}{!}{\includegraphics{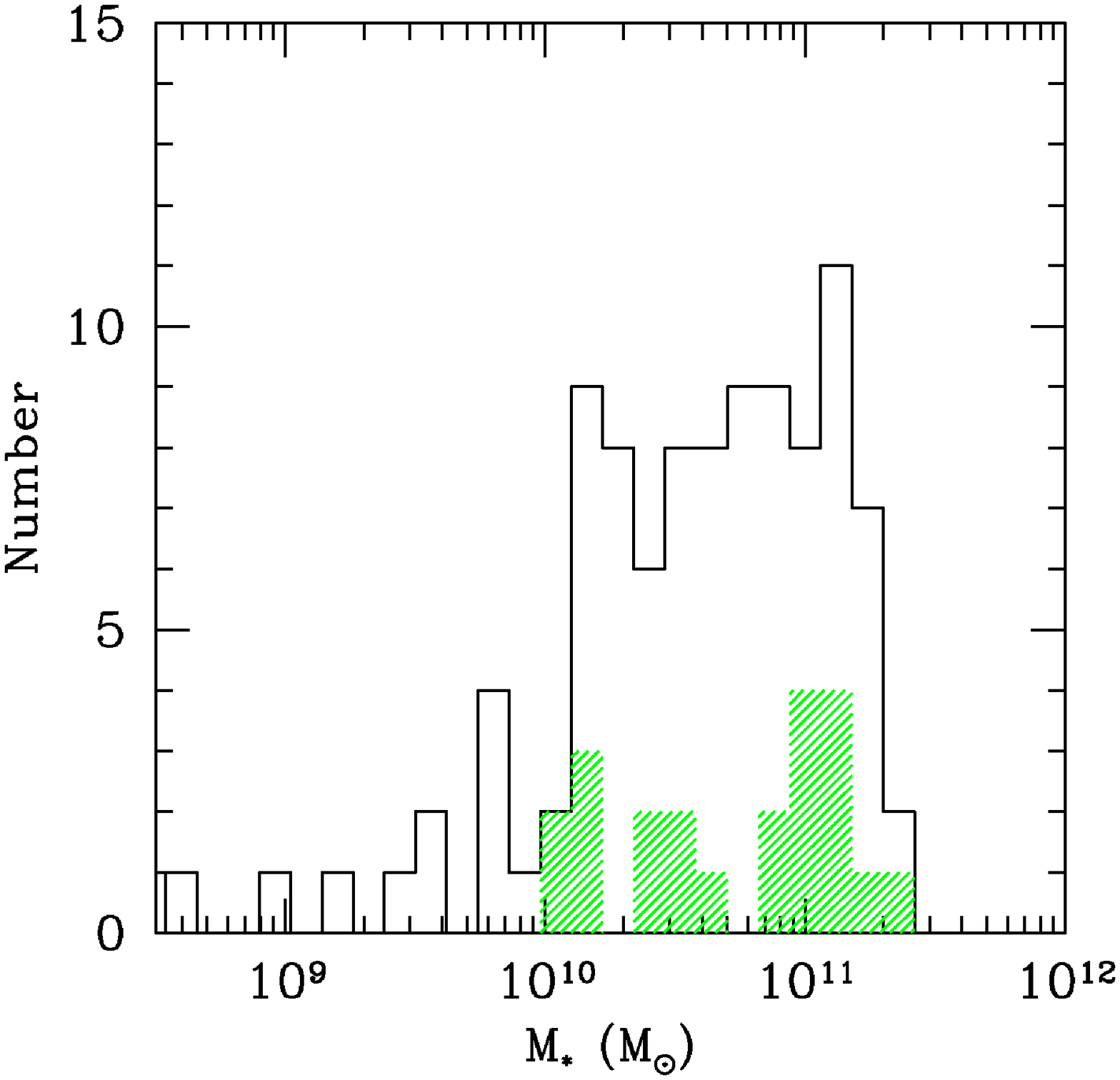}}}
\\
\subfloat{\label{sSFR_hist}\resizebox{43mm}{!}{\includegraphics{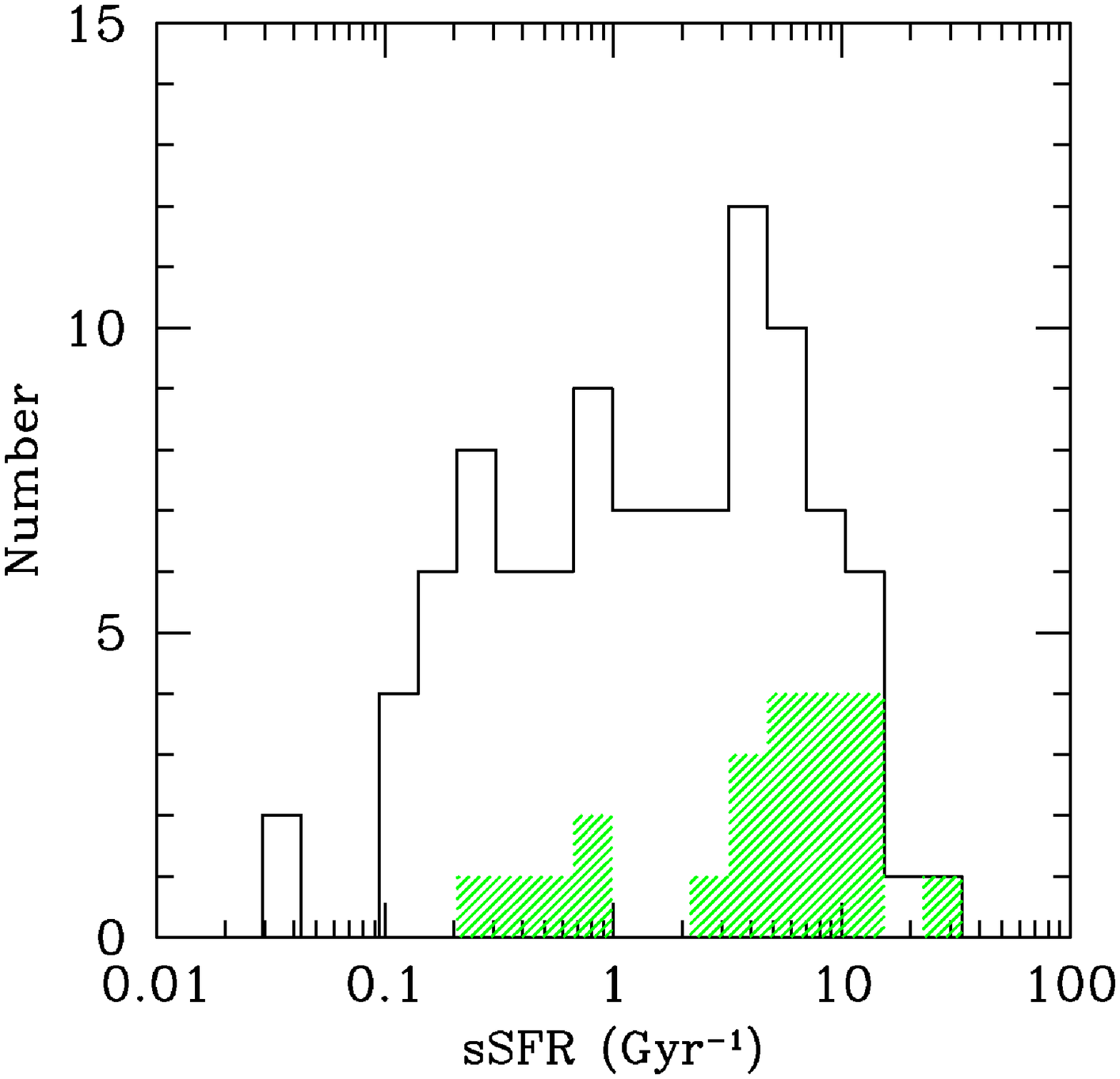}}}
\hspace{2mm}
\subfloat{\label{z_hist}\resizebox{43mm}{!}{\includegraphics{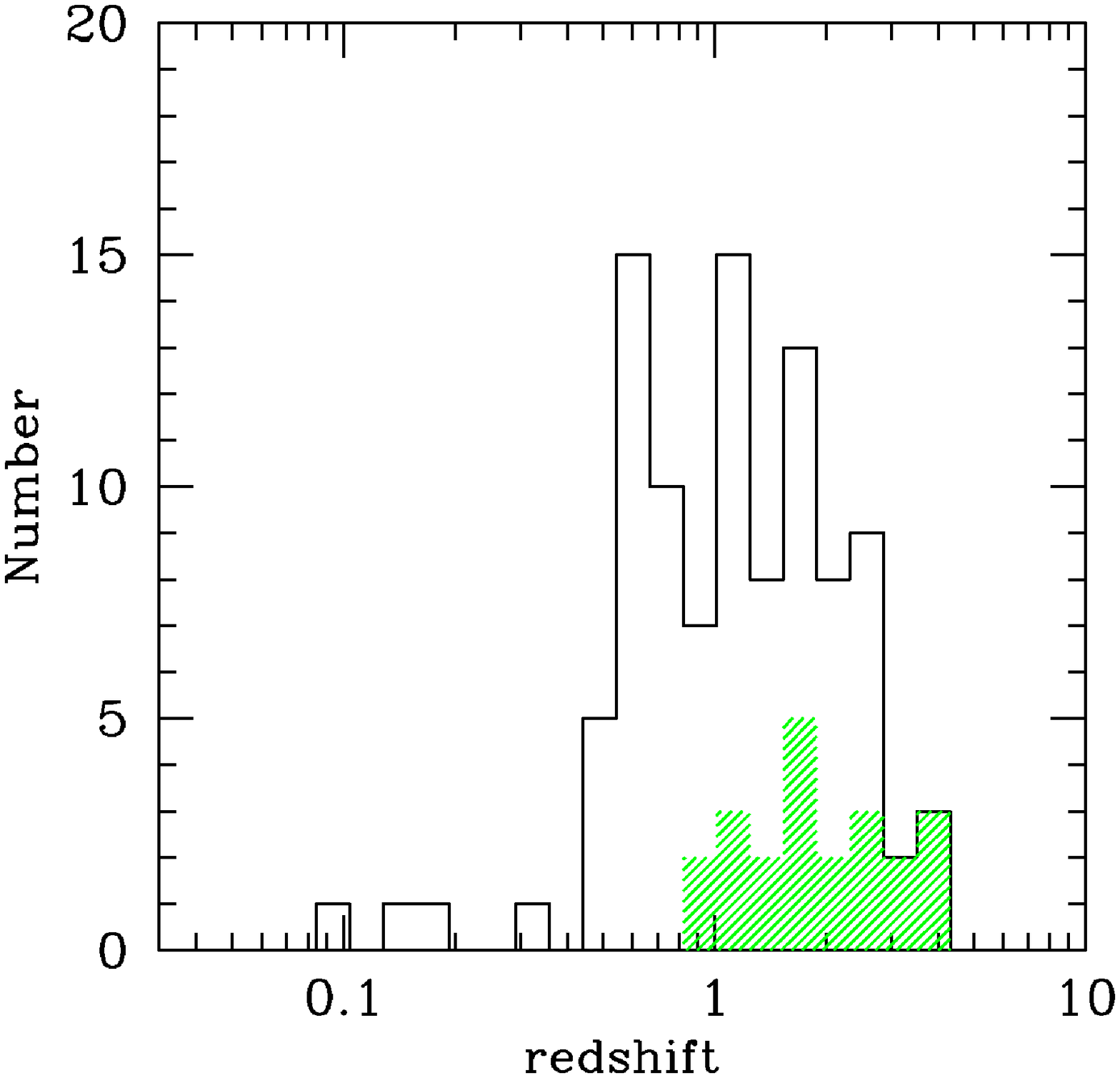}}}
  \caption{Histograms of the basic properties (star-formation rates, stellar masses,
               specific star-formation rates, and redshifts) for the 99 hosts of the
               X-ray AGN sample used in this work. The shaded histograms show the
               same properties for X-ray QSOs, i.e. sources with
               $L_{\rm x}>10^{44}{\rm \,erg\,s^{-1}}$.}
  \label{histograms}
\end{figure}

A fundamental property of each galaxy is its specific star-formation rate (sSFR),
which is defined as the ratio of its star-formation rate to its stellar mass. It is
indicative of how efficient the galaxy is forming stars. To calculate the sSFR for the
AGNs in our sample, we use the star-formation rates measured from the infrared
luminosity. We have calculated the sSFR for 99 X-ray AGNs, 77 with spectroscopic
redshift and 22 with photometric redshift. The SFR, stellar mass, sSFR and redshift
histograms are shown in Fig.\ \ref{histograms}.

\begin{figure}
  \resizebox{\hsize}{!}{\includegraphics{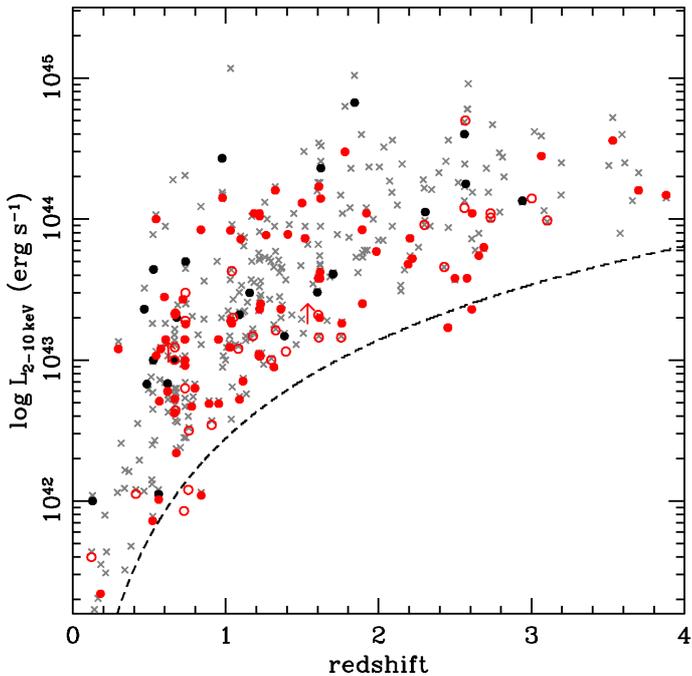}}
  \caption{2--10\,keV luminosities against redshift for all the X-ray sources with
                 redshift determinations in the 3\,Ms \emph{XMM-Newton} survey. The
                 original parent sample is plotted in grey crosses, while black and red
                 filled circles are plotted for the X-ray AGNs with both a stellar mass and a
                 SFR measurement (``broad'' sample). Red symbols represent sources in
                 the GOODS-\emph{Herschel}-PEP area, filled for \emph{Herschel}
                 detections and open for FIR upper limits (``complete'' sample). The line is
                 the $\rm6.5\times10^{-16}\,erg\,s^{-1}\,cm^{-2}$ flux limit, assuming 
                 $\Gamma=1.7$.}
  \label{Lx_z}
\end{figure}

The sample described in the previous paragraph (hereafter the ``broad'' sample) is
not a complete sample of X-ray detected AGNs, because of the various selections of
the sources, which limit their number from 356 to 99. This incompleteness might
affect the statistical properties. In order to account for that, we create a more
complete sub-sample constrained in the \emph{Herschel} area where we have the
most sensitive far-IR measurements. In this area (marked with the small rectangle in
Fig.\ \ref{regions}) for  the X-ray sources for which a far-IR counterpart is not found,
a FIR upper limit is calculated from the sensitivity map of the
GOODS-\emph{Herschel}--PEP survey. Thus, we have 155 X-ray sources, 94 of them
are detected in the FIR, and for 41 we can calculate an upper limit to their
$\rm 100\,\mu m$ and $\rm 160\,\mu m$ fluxes. Twenty sources lie within a
10\,arcsec region of a nearby bright FIR source and an upper limit cannot be
calculated, they are however a random sub-sample, not affecting the completeness.
Out of the 135 (155-20) sources, eight are associated with normal galaxies not
hosting an AGN, a further 12 do not have sufficient optical or mid-infrared
information, or a redshift estimate for an SED fit, and for a further seven the
emission from the AGN dominates over the FIR flux (the AGN-related flux go the
highest wavelength data-point is higher than the star-formation related), making a
star-formation measurement not reliable. Summing up, we calculated the SFRs and
stellar masses of 108 out of the 127 ($\approx85\%$) X-ray AGNs in the
GOODS-\emph{Herschel} region, for which a FIR flux determination is possible (76
detections and 32 upper limits). Hereafter we will call this the ``complete'' sample.
The basic properties (2--10\,keV luminosities and redshifts) of the complete sample
are shown in Fig.\ \ref{Lx_z} with red symbols (filled and open circles for FIR
detections and limits, respectively), while the properties of the overall sample are
plotted in black symbols, and the rest of the X-ray sources are plotted in grey
crosses.

\section{Results}

\subsection{sSFR-$L_{\rm x}$}
\label{sSFR_Lx_result}

\begin{figure}
  \resizebox{\hsize}{!}{\includegraphics{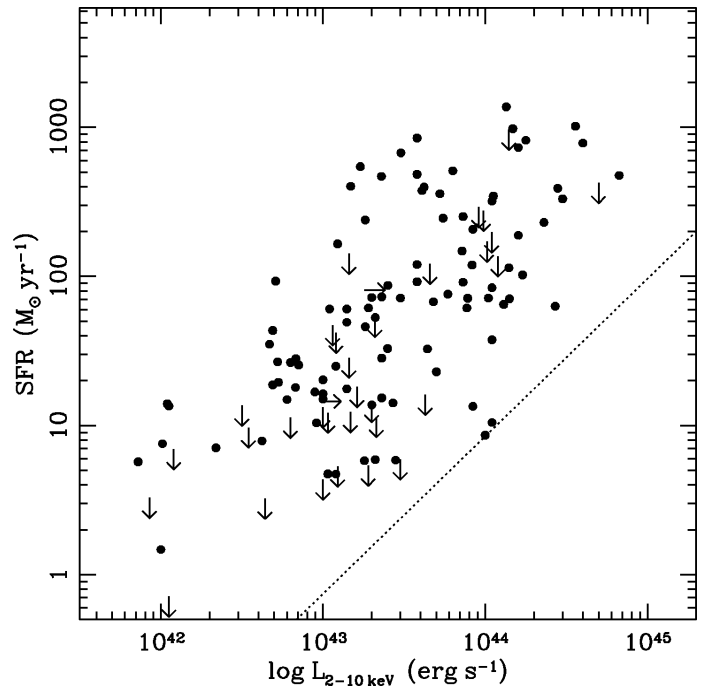}}
  \caption{Star-formation rate against hard X-ray luminosity for X-ray selected
                AGNs. We plot values of 99 far-infrared detected X-ray AGNs, and 32 FIR
                upper limits in the area covered by the deep \emph{Herschel} survey. The
                dotted line is the expected FIR luminosity of a pure-AGN source, translated
                into SFR \citep[see][]{Mullaney2011}. We find a strong correlation between
                the star-formation rate and the X-ray luminosity.}
  \label{SFR_Lx}
\end{figure}

\begin{figure}
  \resizebox{\hsize}{!}{\includegraphics{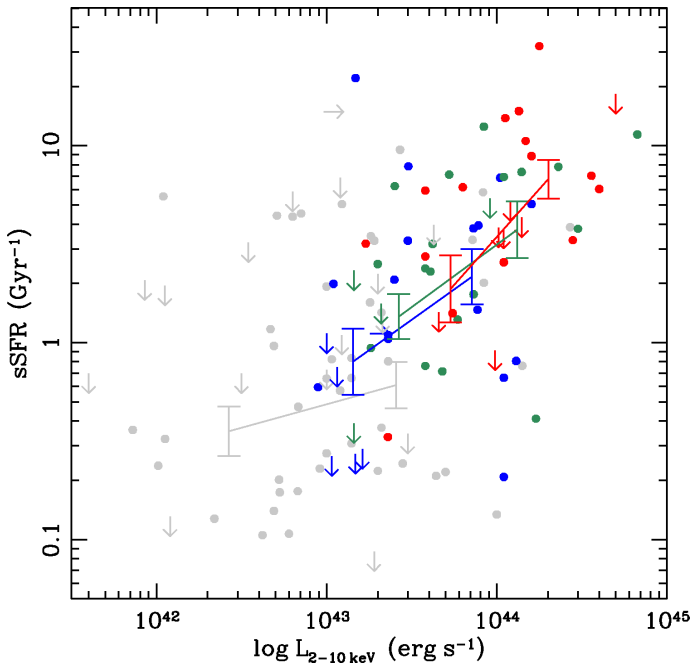}}
  \caption{Specific star-formation rate against hard X-ray luminosity for X-ray
                 selected AGNs. The grey, blue, green and red symbols refer to $z<1.120$,
                 $1.120<z<1.615$, $1.615<z<2.455$ and $z>2.455$, respectively. The
                 error-bars and their associated lines refer to the mean luminosities and
                 specific SFRs of the high- and low-luminosity bins within each redshift
                 bin, using the Kaplan-Meier estimator (see Sect.\ \ref{sSFR_Lx_result}). We
                 do not detect a significant correlation between the X-ray luminosity and
                 sSFR for the lowest redshift bin, but do detect a significant correlation for
higher redshifts ($z\gtrsim1$).}
  \label{sSFR_Lx}
\end{figure}

In previous studies there has been a controversy about the existence of an
observational connection between the AGN and the host galaxy activity. In
Fig.\ \ref{SFR_Lx} we plot the SFR against the hard X-ray luminosity of the 99 X-ray
AGNs with an estimate of the SFR described in the previous section and the 32 upper
limits. There are also two X-ray sources with lower limits in their X-ray luminosities.
These are Compton-thick sources whose X-ray spectra are dominated by a reflection
component according to the spectral fits, and their unobscured luminosities cannot
be determined. The lower limits in Fig\ \ref{sSFR_Lx} are their observed luminosities.
Because of the limits, for the statistical analysis we use the ASURV package
\citep[Rev.\ 1.3][]{LaValley1992}, which implements the methods presented in
\citet{Feigelson1985} and \citet{Isobe1986}. Using the generalised Kendall's $\tau$
method in order to include upper limits, and all the data-points of Fig.\ \ref{SFR_Lx},
we find that the SFR is strongly correlated with the hard X-ray luminosity, with a null
hypothesis probability lower than 0.01\%. To simulate a mass-matched sample and
study the activity of the host independent of its size, we calculate the specific SFRs
of the sample and plot it against the X-ray luminosity in Figure\ \ref{sSFR_Lx}.
Performing the same method, we find again that the two values are strongly
correlated. However, \citet{Mullaney2012} have shown that the correlation between
the X-ray and the infrared luminosity is sensitive on the evolution of the infrared
luminosity with redshift, and this might be affecting the sSFR-$L_{\rm x}$ correlation
we observe here. To test this hypothesis we apply the partial correlation test of
\citet{Akritas1996} and find significant correlations of the sSFR with both redshift
and X-ray luminosities at the $4.6\sigma$ and $4.4\sigma$ levels, respectively.
However, this partial correlation test tends to incorrectly reject the null-hypothesis
in cases where the two ``independent'' parameters (here $L_{\rm x}$ and $z$) are also
correlated with each other \citep[see][]{Kelly2007} as in this case, which limits the
reliability of the test.

In order to further check the effect of the redshift in the sSFR-$L_{\rm x}$ correlation,
we divide our sample into six redshift bins, containing a roughly equal number of
data-points (21 or 22), and repeat the Kendall's $\tau$ method for each bin
separately and for different combinations. The description of the redshift bins and
the results of the test (null hypothesis probability) are shown in
Table\ \ref{statistics}. We can see that there is no correlation at lower redshifts, but
there is a possible correlation for $z>1.15$ (bins 4, 5, 6) with $\gtrsim95\%$
significance. If we merge adjacent bins in order to increase the number of
data-points in each bin, the correlation is again found for $z>1.12$ (bins 4-5, 5-6)
with $>99.8\%$ significance. In this case however the redshift dependence is not
negligible. We also note that within the redshift bins the sample is almost
luminosity-limited. To further check if the X-ray flux limit affects the previous
result, we exclude sources with X-ray luminosities lower then the luminosity limit of
the highest redshift limit of each redshift bin, to create truly luminosity-limited
sub-samples. Repeating the analysis in those sub-samples, the previous result does
not change significantly, except in the highest-redshift bin ($z>2.305$). In
Fig.\ \ref{sSFR_Lx} we colour-code the data-points with respect to their redshifts, in
grey we plot bins 1-2-3, in blue bin 4, in green bin 5 and in red bins 6; the results of
this binning are shown in the last column of Table\ \ref{statistics}. We divide each
data compilation into a low-luminosity and a high-luminosity bin including an equal
number of sources, and plot the mean sSFR and its associated error, calculated using
the Kaplan-Meier estimator in ASURV, and the mean luminosity of the bin. The
sSFR-$L_{\rm x}$ correlation is evident for the blue, green and red data-points
($z>1.15$).

A possibly important factor affecting the previous analysis is the FIR selection of the
sources of our final ``broad'' sample, which reduces the number of X-ray sources
from 356 to 99, being biased in favour of sources with higher SFRs. In order to check
whether this has an effect on the apparent sSFR-$L_{\rm x}$ correlation, we repeat the
previous analysis in the small area covered by the GOODS-{\it Herschel} survey (the
``complete sample''; see Figs.\ \ref{regions} and \ref{Lx_z}). In this case, the
analysis is performed in broader redshift bins due to the smaller number of sources,
and the results are similar to those described in the previous paragraph; there is no
sign of a sSFR-$L_{\rm x}$ correlation below $z\lesssim1.2$, but above this redshift
the correlation is $>95\%$ significant.

We note here that there is a small number of X-ray sources, which are detected in the
far-infrared in a rest-frame wavelength $\rm \lambda>20\,\mu m$, but its flux
density is dominated by the AGN emission, according to the SED decomposition
performed. There are 14 such cases in the ``broad'' sample and seven in the
``complete'' sample with redshifts $z\sim1-3$. These sources could populate the
low-(s)SFR - high-$L_{\rm x}$ area, however their far-IR luminosities cannot be
constrained, not even with an upper limit. If we consider the SFR estimations from
the optical SEDs, although unreliable, they are consistent with the (s)SFR-$L_{\rm x}$
correlation. Moreover, their number is $\sim10\%$ of the sample used, so we are
confident that they will not affect the result.

\begin{table*}
\caption{Results of Kendall's $\tau$ method for the correlation between the specific
               star-formation rate and the hard X-ray luminosity for different redshift
               bins. The null hypothesis probability in each redshift bin is shown in
               column 3 and in combinations of bins in columns 4 and 5, with a lower null
               hypothesis probability meaning a tighter correlation.}
\label{statistics}
\centering
\begin{tabular}{cccccc}
\hline\hline
Bin & number of sources & redshift range & Null Hypothesis (\%) & Null Hypothesis (\%) & Null Hypothesis (\%) \\
\hline
1 & 21 & $0.000-0.620$ & 34    & \multirow{2}{*}{7.9  (bins 1--2)} & \multirow{3}{*}{1.5  (bins 1--3)} \\
2 & 22 & $0.625-0.755$ &  8.5  & \multirow{2}{*}{4.3  (bins 2--3)} &                                   \\
3 & 22 & $0.759-1.113$ & 43    & \multirow{2}{*}{9.5  (bins 3--4)} &                                   \\
4 & 22 & $1.156-1.599$ &  5.4  & \multirow{2}{*}{0.18 (bins 4--5)} & 5.4 (bin 4)                       \\
5 & 22 & $1.605-2.299$ &  0.97 & \multirow{2}{*}{0.11 (bins 5--6)} & 0.97 (bin5)                       \\
6 & 22 & $>2.305$      &  5.2  &                                   & 5.2 (bin 6)                       \\
\hline
\end{tabular}
\end{table*}

\subsection{Redshift evolution}
\label{zevol}

\begin{figure*}
\subfloat[specific SFR]{\label{sSFR_z}\resizebox{90mm}{!}{\includegraphics{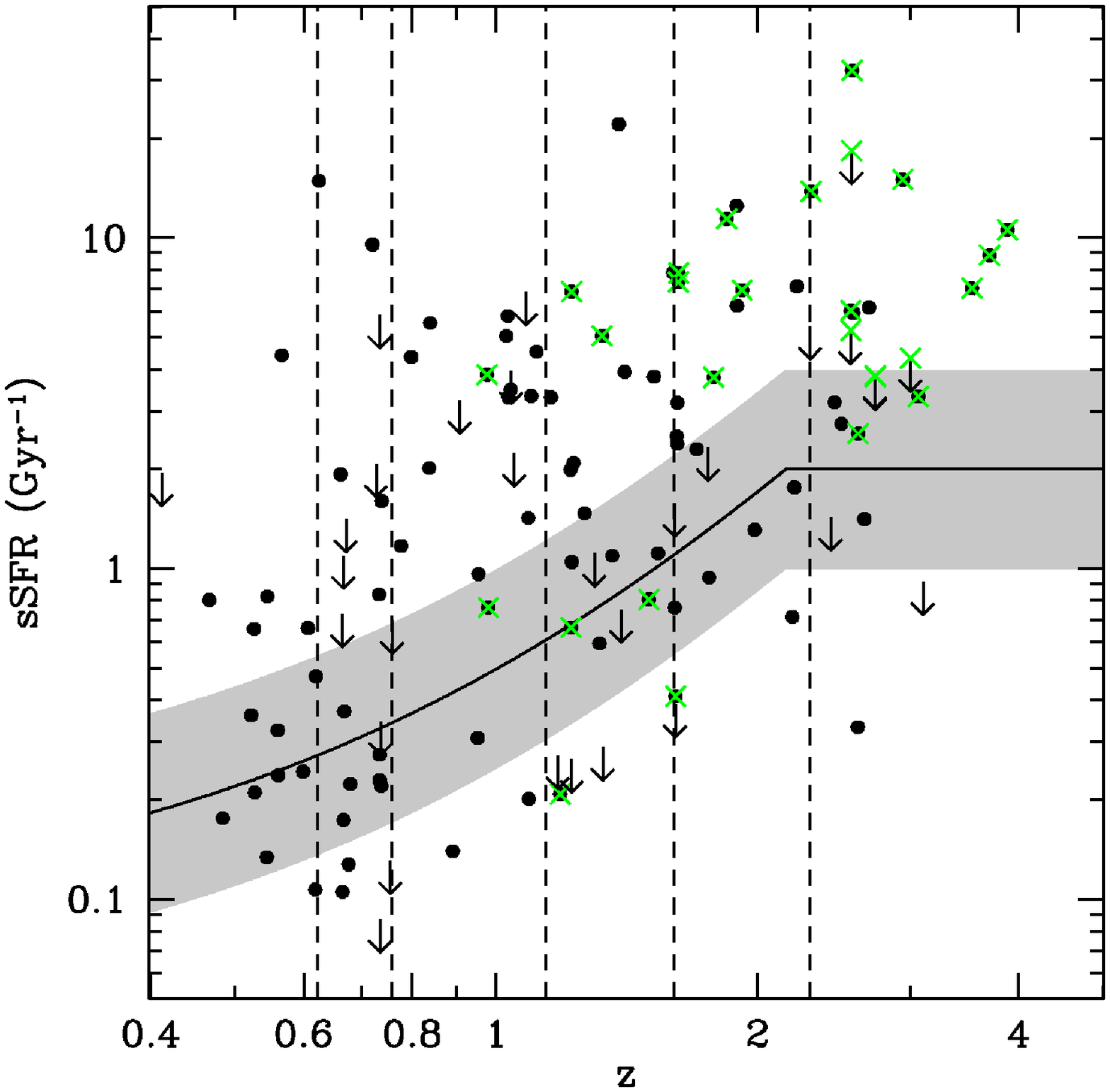}}}
\hspace{4mm}
\subfloat[``starburstiness'']{\label{SB_z}\resizebox{90mm}{!}{\includegraphics{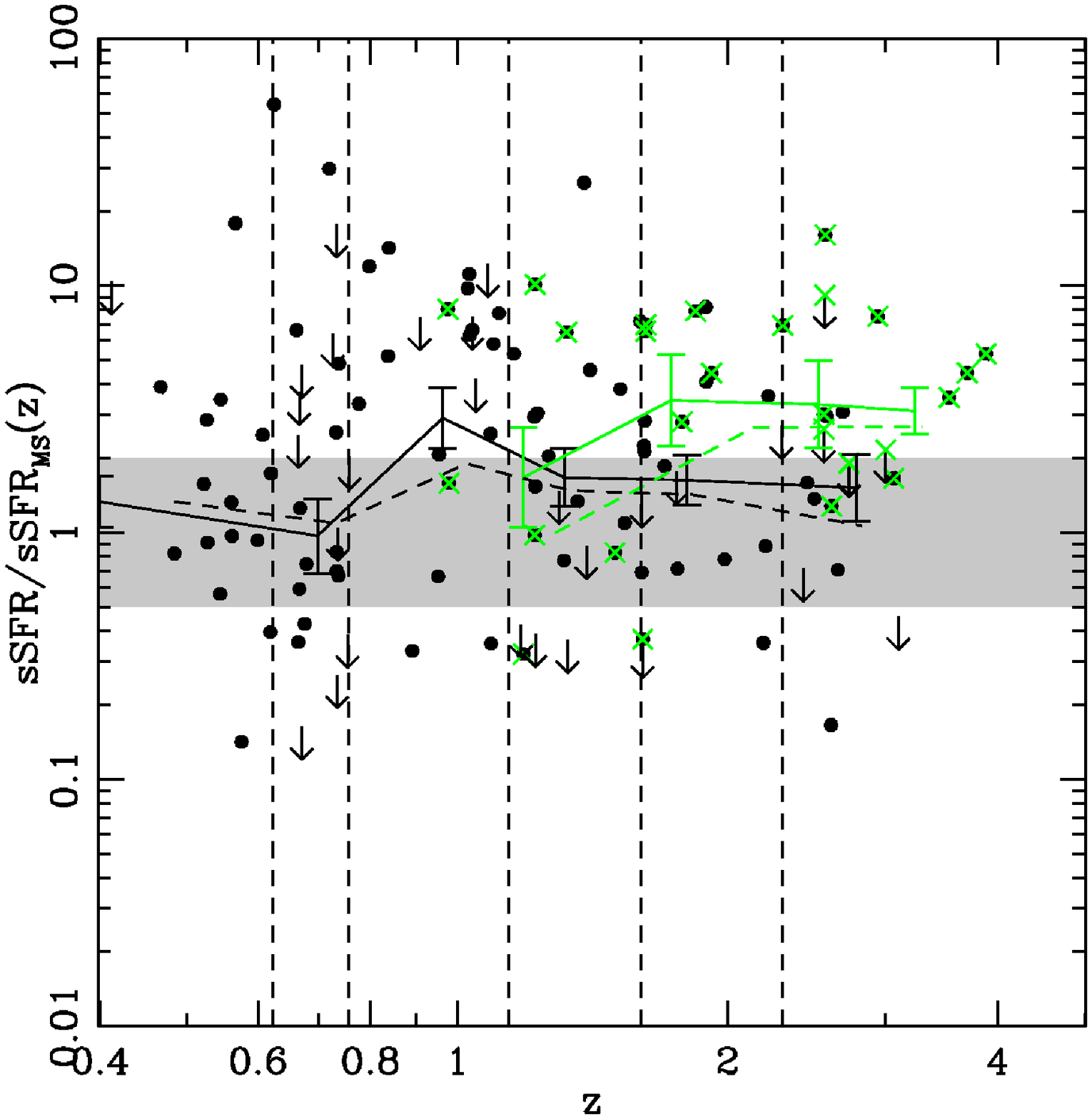}}}
  \caption{Evolution of the specific SFR and the ``starburstiness'' with redshift for
                 X-ray selected AGNs with long wavelength information. The dashed
                 vertical lines refer to the redshift bins of Table\ \ref{statistics}, the solid
                 line in panel (a) is the expected main sequence sSFR according to
                 Equation\ \ref{MS_eqn}, and the grey areas denote the limits of the
                 starburst and quiescent areas. Their borders are double and half the
                 main-sequence sSFR, according to \citet{Elbaz2011}. Green crosses mark
                 the positions of X-ray QSOs having intrinsic
                 $L_{\rm 2-10\,keV}>10^{44}\,{\rm erg\,s^{-1}}$. In panel (b), the solid lines
                 and respective data-points are the running means of the
                 ``starburstiness'' for X-ray AGNs and QSOs in the ``broad'' AGN sample,
                 and the dashed lines are the running means of the ``complete'' X-ray AGN
                 sample (see Sect.\ \ref{final_sample}). There is a general trend for the sSFR
                 of AGN hosts to follow the main sequence, so that the median
                 ``starburstiness'' is constant with redshift. The QSO hosts on the other
                 hand have sSFRs which are somewhat higher.}
 \label{logz_plots}
\end{figure*}

The average sSFR of star-forming galaxies increases with redshift at least up to
$z\approx2$ \citep{Elbaz2007,Daddi2007}, and in this section we investigate how
the hosts of an AGN evolve with respect to the general population. In
Fig.\ \ref{sSFR_z} we plot the sSFR of the AGN hosts against the redshift. The vertical
lines refer to the redshift bins of Table\ \ref{statistics}, while the solid curve is the
expected main-sequence sSFR, according to
\begin{equation}
{\rm sSFR_{MS}}[{\rm Gyr^{-1}}]=\left\{
\begin{array}{l l}
26\times t^{-2.2}_{\rm cosmic}, & z<2.156\\
2,                                         & \rm otherwise
\end{array}
\right.
\label{MS_eqn}
\end{equation}
where $t_{\rm cosmic}$ is given in Gyr. The grey area denotes the borders of the
starburst and quiescent areas, defined as double and half the main sequence sSFR,
respectively \citep{Elbaz2011}. We note here that the increase of the main-sequence
sSFR does not continue forever, and in \citet{Elbaz2011} the density of data-points
supporting the above relation dramatically decreases at $z\gtrsim2.5$. There is
evidence that the main-sequence sSFR is constant above $z\approx2$
\citep{Stark2009,Gonzalez2010} with a value of $\rm sSFR_{MS}\approx2\,Gyr^{-1}$.
According to the above relation the value of $\rm sSFR_{MS}=2$ is reached at
$z=2.156$, so above this redshift we assume a constant relation. There is a hint that
the hosts of the AGNs in our sample are mostly in the main-sequence and starburst
regions, while they generally follow the behaviour of the main sequence with
redshift. With green crosses we mark the positions of X-ray QSOs having intrinsic
$L_{\rm 2-10\,keV}>10^{44}\,{\rm erg\,s^{-1}}$. The majority of them (19/25 sources) are
consistent with being in the starburst region with ${\rm sSFR/sSFR_{MS}}(z)>2$.

This behaviour is also evident if we plot the ``starburstiness\footnote{
${\rm starburstiness\equiv sSFR/sSFR_{MS}}(z)$}'' against redshift in Fig.\ \ref{SB_z}.
The ``starburstiness'' is the ratio of the sSFR of the source over the main-sequence
value at the given redshift. The vertical dashed lines in Fig.\ \ref{SB_z} are identical to
those in Fig.\ \ref{sSFR_z}, and the grey area again marks the main sequence. The
symbols of the data-points are identical to Fig.\ \ref{sSFR_z} (black for all the
sources and green for X-ray QSOs). In each redshift bin we also show the average
``starburstiness'' and its associated statistical uncertainty calculated using the
Kaplan-Meier estimator. For the QSO case we have re-binned the data into four
redshift bins to improve the statistics of the sample. The behaviour seems to
depend on redshift: the QSOs in the first redshift bin with $0.976\leq z\leq1.499$
have an average sSFR consistent with that of the overall population, while higher
redshift QSOs are on average more ``starbursty'', having sSFR more than double that
of the main sequence.

To check how much the complex source selection affects those results, we repeat
the previous analysis for the ``complete'' sample, where we have FIR upper limits for
most of the X-ray sources. We use the same technique, and the result is shown with
the dashed lines in Fig.\ \ref{SB_z}. It is consistent within the statistical uncertainty
with that of the ``broad''sample.  For the QSO case however, the difference between
the complete and the broad sample is significant in the first redshift bin, owing to
the scarcity of such objects. The mean sSFR of the QSOs seems to be in the main
sequence for $z\lesssim2$ and in the starburst region for higher redshifts.

\begin{figure}
  \resizebox{\hsize}{!}{\includegraphics{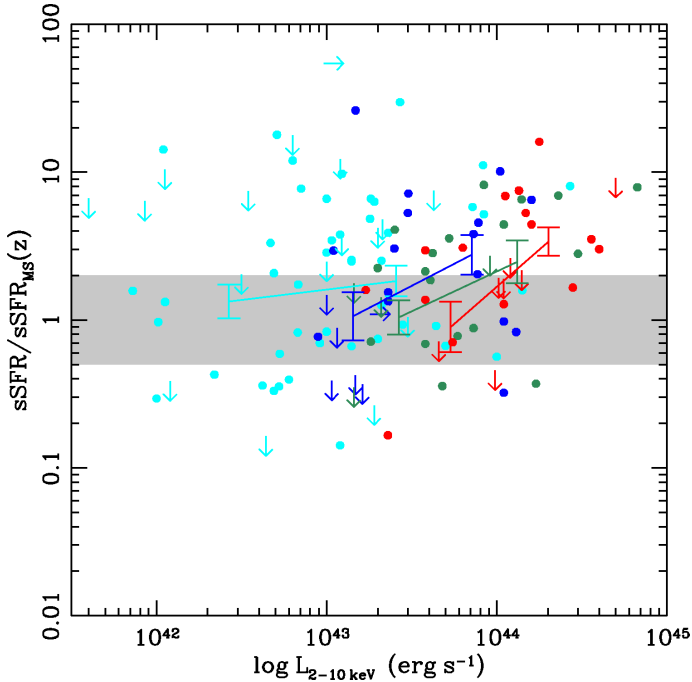}}
  \caption{``Starburstiness'' against (2--10)\,keV X-ray luminosity for the sources
                 in our sample. The grey area is the same as in Fig.\ \ref{logz_plots} and
                 the data-points and error-bars are as in Fig.\ \ref{sSFR_Lx}, substituting
                 the gray point with cyan. We again do not detect any significant
                 correlation between the ``starburstiness'' and the X-ray luminosity for
                 $z\lesssim1$, but there is a correlation for $z\gtrsim1$, within the
                 redshift bins.}
  \label{SB_Lx}
\end{figure}

In order to further investigate how the ``redshift effect'' affects the correlation found
between the sSFR and the hard X-ray luminosity, in Sect.\ \ref{sSFR_Lx_result} we
plot the starburstiness defined in this section against the X-ray luminosity in
Fig.\ \ref{SB_Lx}. The results of the statistical analysis (null hypothesis probability) of
the same redshift bins as in Table\ \ref{statistics} are presented in
Table\ \ref{statistics2}. The analysis shows again a $\gtrsim95\%$ correlation for
redshifts $z\gtrsim1$ and no correlation at lower redshifts. We therefore assume
that the correlation between the host and galaxy activity is not affected by the
evolution of the infrared luminosity with redshift.

\begin{table*}
\caption{Results of Kendall's $\tau$ method for the correlation between the
               starburstiness and the hard X-ray luminosity for different redshift bins. The
               null hypothesis probability in each redshift bin is shown in column 3 and in
               combinations of bins in columns 4 and 5.}
\label{statistics2}
\centering
\begin{tabular}{cccccc}
\hline\hline
Bin & number of sources & redshift range & Null Hypothesis (\%) & Null Hypothesis (\%) & Null Hypothesis (\%) \\
\hline
1 & 21 & $0.000-0.620$ & 84    & \multirow{2}{*}{31   (bins 1--2)} & \multirow{3}{*}{4.4 (bins 1--3)} \\
2 & 22 & $0.625-0.755$ &  6.7  & \multirow{2}{*}{4.5  (bins 2--3)} &                                  \\
3 & 22 & $0.759-1.113$ & 69    & \multirow{2}{*}{29   (bins 3--4)} &                                  \\
4 & 22 & $1.156-1.599$ &  6.0  & \multirow{2}{*}{ 0.5 (bins 4--5)} & 6.0 (bin 4)                      \\
5 & 22 & $1.605-2.999$ &  1.0  & \multirow{2}{*}{ 3.7 (bins 5--6)} & 1.0 (bin 5)                      \\
6 & 22 & $>2.305$      &  5.2  &                                   & 5.2 (bin 6)                      \\
\hline
\end{tabular}
\end{table*}

\subsection{sSFR-$N_{\rm H}$}

\begin{figure}
\subfloat[SFR]{\label{SFR_NH}\resizebox{88mm}{!}{\includegraphics{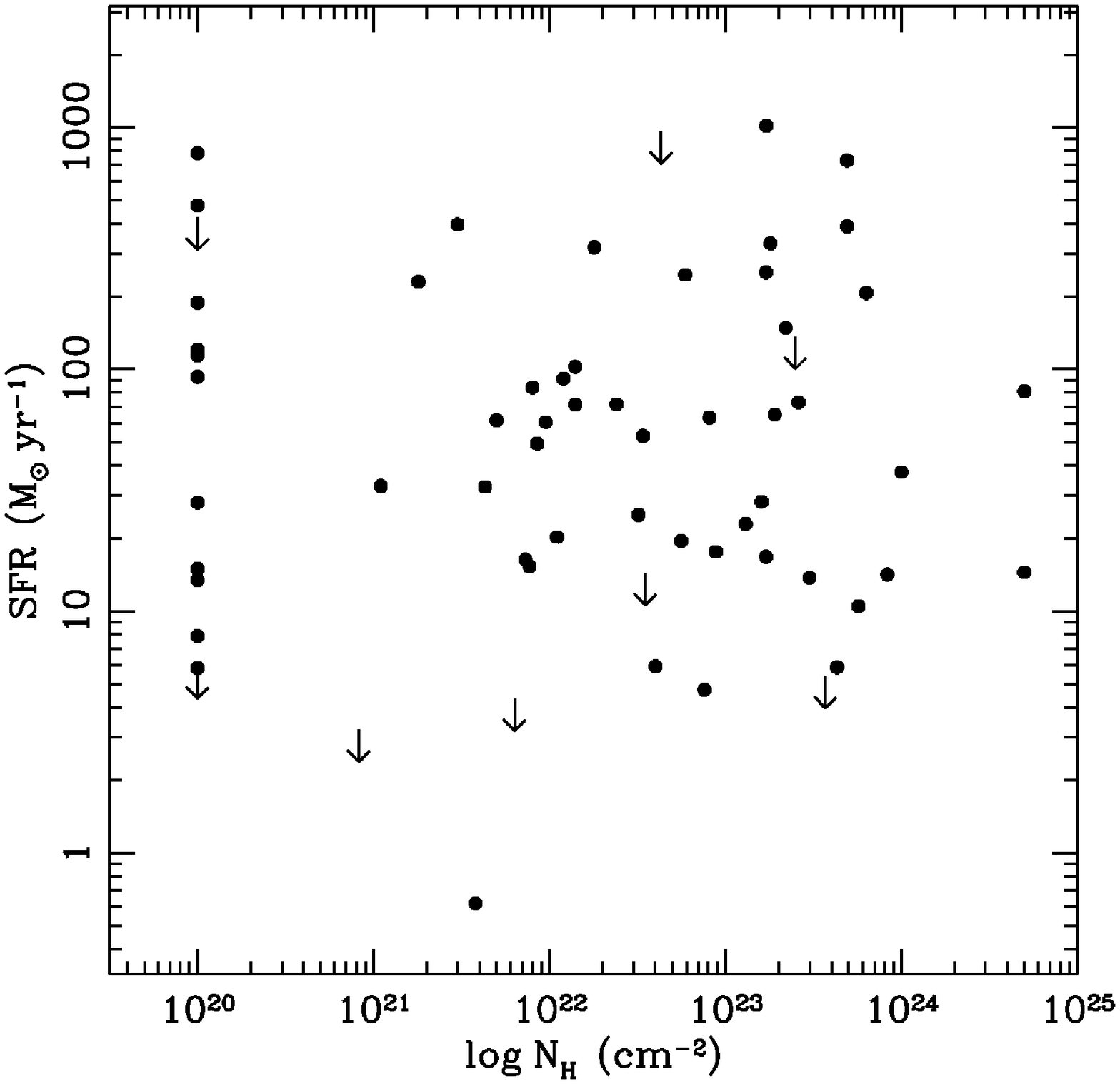}}}
\\
\subfloat[specific SFR]{\label{sSFR_NH}\resizebox{43mm}{!}{\includegraphics{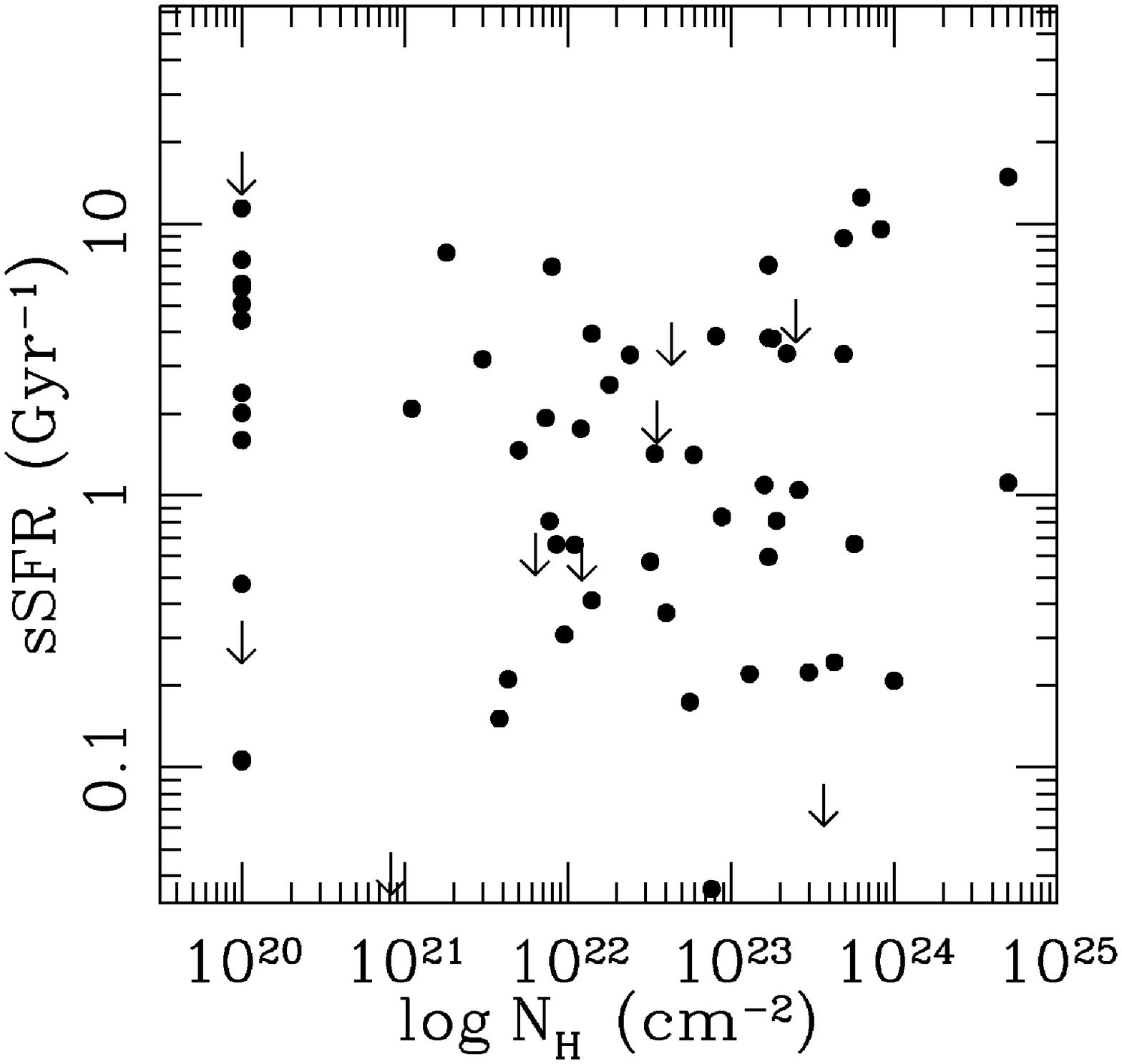}}}
\hspace{2mm}
\subfloat[``starburstiness'']{\label{SB_NH}\resizebox{43mm}{!}{\includegraphics{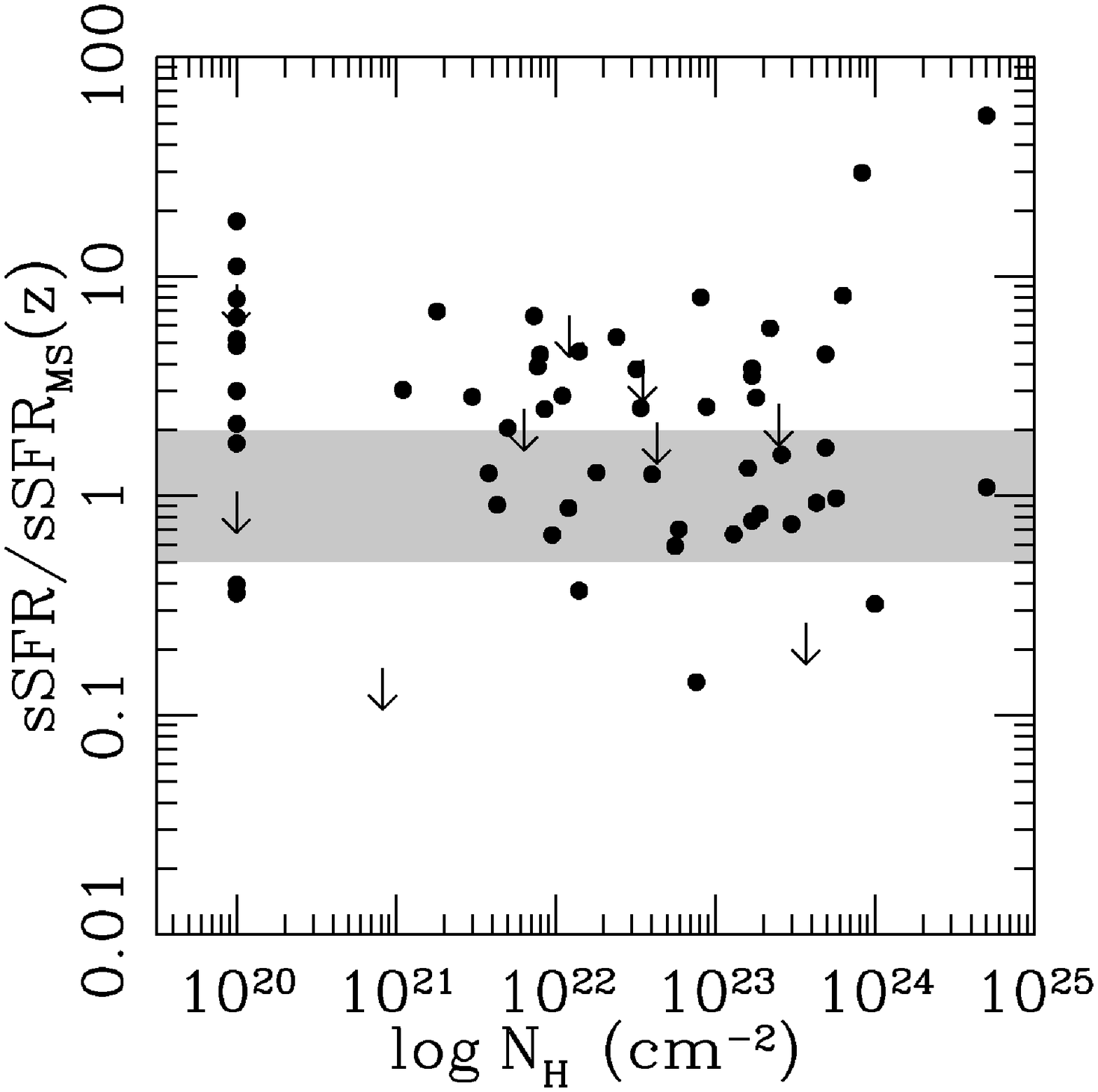}}}

  \caption{Star-formation rate, specific star-formation rate, and ``starburstiness''
                plotted against the hydrogen column density for 65 sources with
                {\it XMM-Newton} spectra and robust SED fitting (including nine with
                \emph{Herschel} upper limits). The grey area in panel (c) is the
                main-sequence, as in Figs.\ \ref{logz_plots} and \ref{SB_Lx}. The errors on
                the hydrogen column density are in the range of $\lesssim10\%$ to 30\%.
                We do not detect any significant correlation between the two values, even
                if we split the sample into redshift or luminosity bins, in any of the three
plots.}
  \label{NH_plots}
\end{figure}

In earlier studies there have been some hints of a correlation between the
star-forming activity of the host and the AGN obscuration in the X-rays.
\citet{Page2004} found that X-ray absorbed sources are more likely to be detected at
sub-mm wavelengths because of their extreme star-formation rates, although the
absorbed AGN sample consisted only of type\,I (broad-line) QSOs \citep{Page2001},
which are not a representative sample. Moreover, \citet{Alexander2005} found that
the majority of radio-detected SCUBA sub-mm sources are consistent with being
heavily obscured AGNs, with $N_{\rm H}\gtrsim10^{23}\,{\rm cm^{-2}}$, although the
active nucleus is not bolometrically dominant. \citet{Bauer2002} found hints that
X-ray sources with sub-mJy radio counterparts (tracing star formation) are, on
average, more obscured than the unmatched population, confirmed by
\citet{Georgakakis2004}. Subsequently, \citet{Rovilos2007b} using a combination of
the 1\,Ms CDFS and the E-CDFS surveys found that such a trend was confined only
to AGNs with any evidence for X-ray obscuration ($N_{\rm H}>10^{21}\,{\rm cm}^{-2}$),
linking it with line-of-sight effects. However, deeper surveys both in X-rays and at
infrared wavelengths failed to reproduce those results
\citep[e.g.][]{Lutz2010,Rosario2012,Trichas2012}. Here, we use the deepest
{\it XMM-Newton} survey, providing good quality X-ray spectra, combined with the
deepest {\it Herschel} PACS observations, and an SED decomposition technique to
clarify this issue. In Fig.\ \ref{SFR_NH} we plot the hydrogen column density of the
sources for which we have good X-ray spectra against their star-formation rates. We
apply a value of $N_{\rm H}=10^{20}\,{\rm cm}^{-2}$ to sources which show no signs of
obscuration and $N_{\rm H}=5\times10^{24}\,{\rm cm}^{-2}$ to Compton-thick AGNs.
We do not find any significant correlation between the two values. Using the
Kendall's $\tau$ method we find a null-hypothesis probability of 90\%. Excluding
unobscured AGNs or splitting the data into redshift bins does not change this result;
the null hypothesis probability is always higher than 20\%. Neither is there a
significant correlation in the high luminosity AGN
($L_{\rm x}>10^{44}{\rm \,erg\,s^{-1}}$) or the high redshift ($z>1.5$) sub-samples. In
order to simulate a mass-matched sample and correct for any redshift effects in the
column density and sSFR values \citep[e.g.][]{Hasinger2008} we also plot the specific
SFR and the starburstiness against $N_{\rm H}$ in Figs.\ \ref{sSFR_NH} and
\ref{SB_NH}. Performing all the previous tests, we again do not find any significant
correlation, except for the sSFR and starburstiness of AGNs with $0.7<z<1.4$, where
we find hints of an anti-correlation at the 95\% level. However, considering the
behaviour of the overall sample and the complex selection effects to shape that
sub-sample, we do not consider it important. This behaviour is in broad agreement
with models assuming a clumpy absorber \citep{Elitzur2006,Nenkova2008}, where
the absorption strongly depends on the number of absorbing clumps crossing the
line-of-sight, however there is a number of obscured AGNs (with
$N_{\rm H}>10^{22.5}\,{\rm cm}^{-2}$) and ${\rm sSFR/sSFR_{MS}}(z)<1$ which are still
hard to explain with these models.

\subsection{Rest-frame colours}

The colour-magnitude diagram (CMD; the rest-frame $U-B$ or $U-V$ colour plotted
against the absolute $B$ or $V$ magnitude) is used in a number of studies to check
the evolutionary stage of the AGN hosts. The hosts of AGNs are concentrated in and
around the ``green valley''
\citep{Nandra2007,Rovilos2007,Silverman2008,Georgakakis2008,Hickox2009,Georgakakis2011},
which is thought to signpost the transition phase from a starburst to a ``dead''
elliptical. There are however a number of factors that affect the position of a source
(especially an AGN) in the CMD making the meaning of the above observation
unclear. For example, it has been noted that both the AGN contribution and dust
obscuration can alter the observed optical colours of the AGN hosts
\citep{Pierce2010,Cardamone2010b,Lusso2011}, making them bluer or redder.
Moreover, AGNs are usually found in relatively high stellar mass hosts
\citep[$M_{\star}\approx10^{10}-10^{12}\,{\rm M_{\sun}}$;][this study]{Kauffmann2003,Brusa2009,Xue2010,Mullaney2012},
and that in turn makes them ``avoid'' the blue cloud; it is observed that the
concentration of AGNs around the ``green valley'' is not detected when using
mass-matched samples \citep{Silverman2009,Xue2010,Mullaney2012}. In this
section we use the sample of AGNs for which we have an independent way to measure
the star-formation activity, to check the validity of the colour-magnitude diagram
without correcting the optical magnitudes for the host contribution or dust
reddening. In addition, we use a colour-mass diagram instead of the
colour-magnitude approach, in order to simulate a mass-matched sample.

\begin{figure}
  \resizebox{\hsize}{!}{\includegraphics{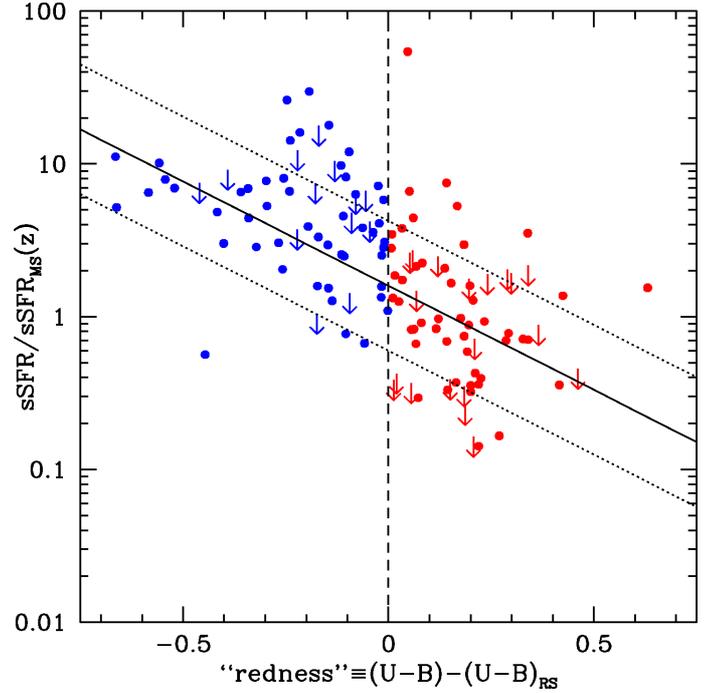}}
  \caption{Starburstiness against deviation from the dividing line between red and
                blue galaxies for the sources of our sample. With blue and red symbols are
                plotted galaxies in the blue cloud and the green valley, and red-sequence
                galaxies respectively; the dividing line (the border of the red sequence) is
                the dashed vertical line, calculated from the parametrisation of
                \citet{Peng2010}. The lines denote the best-fit model and its standard
                deviation, according to Equation\ \ref{SB_RS_eqn}.}
  \label{SB_RS}
\end{figure}

The positions of the red sequence, blue cloud and green valley in the
colour-magnitude (and colour-mass) diagram are strongly dependent on redshift
\citep{Bell2003,Borch2006,Peng2010}. In this section we use the $U-B$ rest-frame
colours of a sample of AGN hosts spanning from $z\sim0.1$ to $z\sim4$ and we
use the parametrisation of \citet{Peng2010} to define the dividing line between blue
and red galaxies, extrapolated to higher redshifts. \citet{Xue2010} have
demonstrated that the colour bi-modality of non-AGNs exists up to at least
$z\approx3$. We use the fitted SEDs including the AGN contribution and the filter
curves of the COMBO-17 survey \citep{Bell2003} to measure the optical colours of
the AGN hosts. In Fig.\ \ref{SB_RS} we plot the starburstiness of the AGN hosts
against their ``redness'', defined as the deviation of their rest-frame colours from
the dividing line in the colour-mass diagram. We observe a clear anti-correlation
between the two values, which is statistically significant at a $>99.9\%$ level. We
parametrise it using the Buckley-James regression method \citep{Buckley1979}:
\begin{equation}
\log\frac{\rm sSFR}{{\rm sSFR_{MS}}(z)}=(-1.36\pm0.17)[(U-B)-(U-B)_{\rm RS}]+(0.2\pm0.4)
\label{SB_RS_eqn}
\end{equation}
(solid and dotted lines in Fig.\ \ref{SB_RS}). In principle, this significant
anti-correlation allows us to use the colour-mass diagram as a diagnostic of the
host properties when detailed observational information, which would allow the
determination of accurate star-formation rates and stellar masses, is not available,
although the large scatter limits its reliability. Moreover, the AGN sample probed here
is only a sub-sample of the total AGN population, selected both in the X-rays and at
longer (far-infrared) wavelengths. There is evidence that the different selections of
AGNs bias their position in the colour-magnitude diagram \citep{Hickox2009}, with
the X-ray-selected AGNs being in the green valley ($(U-B)-(U-B)_{\rm RS}\sim0$ in
Fig.\ \ref{SB_RS}), the infrared-selected in the blue cloud and the radio-selected in
the red sequence. The complex selection of the AGNs in our sample limits its
representativeness.

\section{Discussion}

\subsection{Is there an AGN-host correlation?}

\subsubsection{Low redshifts ($z\lesssim1$)}

Previous studies in the low-redshift Universe \citep{Netzer2009,Serjeant2009} have
shown a correlation between the AGN luminosity (bolometric or optical) and the
host galaxy luminosities (or star-formation rates), probing luminous QSOs with
$L_{\rm x}\gtrsim10^{43}{\rm \,erg\,s^{-1}}$ and optically-selected QSOs, respectively.
On the basis of our low-redshift ($z\lesssim1$) -- low-luminosity
($L_{\rm x}\lesssim10^{43.5}{\rm \,erg\,s^{-1}}$) AGN sample, we do not see such clear
signs of a correlation between the AGN and the host galaxy activity (parametrised by
the sSFR; see Figs.\ \ref{sSFR_Lx} and \ref{SB_Lx} and Tables\ \ref{statistics} and
\ref{statistics2}). This implies that lower luminosity AGN activity, especially at low
redshifts, is not directly linked to the state of the host galaxy, if the latter is
parametrised by its sSFR or its ``starburstiness''. In the redshift range of the first
three bins ($z<1.12$) there are seven quiescent AGN hosts and 11 starbursts, while
the mean sSFR is within the borders of the main sequence. The selection of objects
which have a FIR counterpart could affect the mean ``starburstiness'' of our sample,
but this effect is found to be minimal when limiting the sample to the area where
upper FIR limits are available. It is likely that the AGN process takes place as a result
of instabilities that affect the nucleus but do not have any prominent effect overall, 
being only confined to circumnuclear star formation. Indeed, there is a positive
correlation of the AGN power with the nuclear star formation in local Seyfert-1
galaxies \citep{Thompson2009,DiamondStanic2012}, and such a correlation is also
supported by recent models \citep{Hopkins2010a}. The causal mechanism behind
this connection could be high-mass stellar winds fueling the AGN
\citep[see e.g. simulations by][]{Schartmann2009}. This nuclear correlation however
does not leave a clear mark on the overall observable properties of the system
(infrared and X-ray luminosities).

\subsubsection{Higher redshifts ($z\gtrsim1$)}

In the more distant Universe there are studies finding both a correlation between the
AGN power and star-formation intensity
\citep{Trichas2009,Hatziminaoglou2010,Bonfield2011}, and no signs of any
\citep{Seymour2011,Rosario2012}, using different diagnostics and source selections,
while there is evidence that the star-formation rates of the AGN hosts are enhanced
with respect to those at $z\lesssim1$ \citep{Mullaney2010}. Combining data at
different luminosities and redshifts, \citet{Lutz2010} and \citet{Shao2010} propose
different mechanisms for the fuelling of both the star formation and the AGN,
merger-driven for high luminosities and secular for lower, with the
``high-luminosity'' limit being strongly dependent on redshift
\citep[see also][]{Serjeant2009,Wilman2010}. Recently, \citet{Mullaney2012} using
deep {\it Chandra} and {\it Herschel} observations find that the increase of both
infrared and X-ray luminosities with redshift affect the observed
$ L_{\rm x}-{\rm SFR}$ correlation, and do not detect it for moderate luminosity AGNs
($L_{\rm x}=10^{42}-10^{44}\,{\rm erg\,s^{-1}}$) in individual redshift bins. However,
such a correlation emerges at redshifts $z\sim1-2$, if the stacked signal from
individually X-ray undetected AGNs is factored in, revealing a similar
$L_{\rm x}-M_{\star}$ relation to the SFR main sequence \citep{Mullaney2012b}. There is
also evidence that the correlation is weaker at higher redshifts for the highest
luminosity AGNs \citep{Rosario2012}. 

Here, we use a combination of the deepest {\it XMM-Newton} and {\it Herschel}
observations in combination with an SED decomposition technique to remove any
AGN flux from the far-IR wavelengths \citep[see][]{Mullaney2011}, and find a
correlation between the specific SFR and the X-ray luminosity for $z\gtrsim1$ and
$L_{\rm x}\gtrsim10^{43}{\rm \,erg\,s^{-1}}$. \citet{Mullaney2012} using a similar
sample with a somewhat lower luminosity range
($L_{\rm x}=10^{42}-10^{44}{\rm \,erg\,s^{-1}}$) fail to detect a significant correlation at
those redshifts, suggesting that the higher luminosity sources are responsible of the
correlation detected in this work. Our sample is highly incomplete for
$L_{\rm x}<10^{43}{\rm \,erg\,s^{-1}}$ at $z\gtrsim1$ (see Fig.\ \ref{Lx_z}). Following
the discussion of the previous section, this behaviour can be either because the
nuclear star formation in high X-ray luminosity objects is so strong that it dominates
over that of the host, or there is a link between the AGN activity and the evolution of
the host galaxy at higher redshifts and luminosities, parametrised by the overall
sSFR. In the former case we would expect a correlation between the obscuration of
the AGN and the star-formation rate \citep[see e.g.][]{Ballantyne2006}, at least for
high luminosity objects showing some degree of obscuration, because of the
expected increase in the covering factor and the column density of the obscuring
material, if it is also responsible for the star formation. In Figs.\ \ref{sSFR_NH} and
\ref{SB_NH} we do not detect any such correlation; moreover, according to
\citet{Ballantyne2008}, a circumnuclear star-forming disk could not sustain very
high star-formation rates ($\rm\gtrsim10\,M_{\sun}\,yr^{-1}$, typical of rates of our
sample) and is not compatible with a high-luminosity AGN, because it would limit the
necessary gas supply. This is an indication that the star-formation rate is not nuclear
and therefore not directly connected to the AGN obscuration. The lack of any
correlation between the AGN obscuration and the sSFR also indicates that the
star-forming gas is not directly connected to the AGN obscuration, so the
obscuration from the host galaxy \citep[see][]{MartinezSansigre2009} is not
dominant. The apparent connection between the galaxy and AGN activity is therefore
likely evolutionary.

This correlation of star formation at galaxy scales with the AGN activity seems to be
in disagreement with models suggesting that the AGN outflows quench the
star-forming activity by disrupting the cold gas supply \citep{DiMatteo2008}, as
there are a number of AGNs with high luminosities,
$L_{\rm 2-10\,keV}>10^{44}\,{\rm erg\,s^{-1}}$, which are actively star-forming, with
$\rm sSFR>1\,Gyr^{-1}$, and the most active AGNs appear to be more ``starbursty''
than lower $L_{\rm 2-10\,keV}$ sources, at least in the redshift range
$2\lesssim z\lesssim3$ (see Figs.\ \ref{Lx_z} and \ref{SB_z}). Such behaviour is
consistent with the suggestion that the AGN activity might enhance the star-forming
activity of the host galaxy instead of quenching it \citep[see e.g.][]{Elbaz2009} and
one means of doing that is through the disruption of the density profile of the host
by an AGN-generated jet \citep[see also][]{Gaibler2011}. An issue that has to be
addressed in this case is whether the jet would be detected at radio wavelengths,
since only 2/7 of the highest sSFR and highest $L_{\rm 2-10\,keV}$ sources in
Fig.\ \ref{sSFR_Lx} are detected in the radio, one of them being only marginally
radio-loud with $L_{\rm 1.4\,GHz}=6\times10^{31}\,{\rm erg\,s^{-1}\,Hz^{-1}}$. We note
that the radio luminosity of $\rm HE\,0450-2958$
\citep[the source studied in][]{Elbaz2009} is not radio-loud according to our
classification, having a radio luminosity in the 1.4--GHz band in the order of
$\rm 10^{31}\,erg\,s^{-1}\,Hz^{-1}$ \citep{Feain2007}, meaning that a relatively low
radio luminosity jet could cause a star-formation episode.

\subsection{Where do AGNs live?}

Recent studies \citep{Daddi2007,Daddi2009,Dunne2009,Pannella2009,Magdis2010,Elbaz2011}
have found a relation between the star-formation rate and the stellar mass,
consistent with being linear at all redshifts from local to $z\sim4$, but where the
normalisation of this relation is strongly dependent on redshift
\citep{Karim2011,Elbaz2011}. In this discussion we use the star-formation
``main-sequence'' of \citet{Elbaz2011} up to $z=2.156$ and a constant value of
$\rm sSFR_{MS}=2$ thereafter. As we can see in Fig.\ \ref{sSFR_z}, the sSFRs of the
AGN hosts are mostly on the main sequence, indicated by the grey area or above it.
This is more clearly demonstrated in Fig.\ \ref{SB_z}, where we plot the deviation
from the main sequence (``starburstiness'') of the AGN hosts. The black data-points
and line denote the running mean (and the respective error) of the whole AGN sample
described in \S\,\ref{final_sample}; the line is constant with redshift (within the
errors) and close to the upper border of the main sequence. This result also holds if
we use a sample unbiased by the lack of upper limits for all the FIR-undetected
sources (dashed line -- see \S\,\ref{zevol}). This is a similar result to \citet{Xue2010}
who find that the SFR of AGN hosts is similar to that of non-AGN galaxies when
using mass-matched samples for $z\lesssim3$.

Overall, there are 11 quiescent, 54 starburst and 34 main-sequence AGN hosts,
which is in agreement with the findings of \citet{Santini2012} who use similar
methods on a wider sample. Within the luminosity range
$10^{42}<L_{\rm x}<{\rm 10^{44}\,erg\,s^{-1}}$ we find 23/69 main-sequence, 38/69
starburst and 8/69 quiescent hosts (assuming an $1\,\sigma$ confidence interval
of a binomial distribution). These numbers do not agree at first glance with the
findings of \citet{Mullaney2012} who use a sample similar to the one used in this
study. However, \citet{Mullaney2012} use a wider main-sequence region (a factor of
three instead of a factor of two of the main-sequence sSFR) and if we adopt this
definition, the above numbers become 39/69, 27/69, and 3/69, respectively, much
closer to \citet{Mullaney2012}. The residual difference of the fewer quiescent hosts
found here is because of the stacking analysis done in \citet{Mullaney2012} to
estimate the behaviour of FIR undetected AGNs. The limited number of sources in the
``complete'' sample in the area covered by {\it Herschel}-PACS does not allow us to
perform such an analysis here. We note that \citet{Santini2012} find similar results
when they factor-in their stacking analysis of undetected AGNs. The increased mean
sSFR of the AGN hosts we find in this study is in line with the $L_{\rm x}-{\rm sSFR}$
correlation, suggesting that the AGN and star-formation processes are connected,
either affecting each other, or having a common cause. The most luminous AGNs
with $L_{\rm x}>10^{44}{\rm \,erg\,s^{-1}}$ (X-ray QSOs) are represented with green
symbols in Figs.\ \ref{sSFR_z} and \ref{SB_z}, and reside on average in the region of
starburst galaxies (defined as having ${\rm sSFR/sSFR_{MS}}(z)>2$) for $z\gtrsim2$,
which reflects the overall correlation between the AGN luminosity and the host
activity.

In the redshift range $1<z<2$ there are a few high-luminosity AGNs which have very
low sSFR and ``starburstiness'' values, placing them in the main sequence or even in
the quiescent region. Although these objects are not enough to disrupt the
sSFR-$L_{\rm x}$ correlation at those redshifts (see Figure\ \ref{sSFR_Lx}), they could
be examples of the powerful AGN suppressing the star-formation. In a recent study,
using \emph{Chandra} X-ray data and \emph{Herschel}-SPIRE sub-mm
($\rm 250\,\mu m$) data in the CDF-N, \citet{Page2012} find that the highest X-ray
luminosity ($L_{\rm x}>10^{44}\,{\rm erg\,s^{-1}}$) AGNs are rarely detected in the
sub-mm wavelengths, and therefore have modest SFRs
\citep[see also][]{Trichas2012}. In our sample, most of the X-ray QSOs (14/20 of the
``complete'' sample) are detected in the far-infrared, although at a shorter
wavelength ($\rm 100\,\mu m$) than in the sample of \citet{Page2012}. This could
imply that there is some residual contribution from the AGN in shorter FIR
wavelengths. However, with our SED analysis we identify and remove the contribution
of AGN flux in the far-infrared flux, so this explanation is unlikely.
\citet{Rosario2012} argue that the SFR-$L_{\rm x}$ relation starts to weaken above
$z\approx1$, and indeed the correlation we find is not very strong for the
$1.156<z<1.599$ redshift bin, as a result of the low-sSFR QSOs in that redshift bin.
We do find on the other hand, that at higher redshifts the sSFR-$L_{\rm x}$ correlation
is stronger, and the high X-ray luminosity AGNs are on average more ``starbursty''
than the overall sample. This could be a result of higher abundance of molecular gas
at higher redshift \citep[see e.g.][]{Daddi2010,Bournaud2011}, where despite the
feedback from the powerful AGN, the star-formation is still powerful.

\section{Conclusions}

We select 131 AGNs from the 3\,Ms {\it XMM-Newton} survey and measure their
star-formation rates using long wavelength far-IR and sub-mm fluxes with
rest-frame wavelength above $\rm 20\,\mu m$. For 32 of the 131 sources we are
able to derive only an upper limit of the star-formation rate. We take special care in
modelling the spectral energy distributions, identifying and removing the AGN
contribution, and derive the sSFR and stellar masses of the hosts, comparing them to
the AGN properties (X-ray luminosity and absorption). Our results can be
summarised as follows:
\begin{enumerate}
\item We find no evidence for a correlation between the sSFR and the X-ray
          luminosity for sources with $L_{\rm x}\lesssim10^{43.5}{\rm \,erg\,s^{-1}}$ and at
          $z\lesssim1$.
\item We find a correlation between the sSFR and the X-ray luminosity for sources
          with $L_{\rm x}\gtrsim10^{43}{\rm \,erg\,s^{-1}}$ and at $z\gtrsim1$. There is no
          indication that this correlation is a result of a redshift effect, as it is present
          even when we divide the data into narrow redshift bins. We argue that it is
          instead a result of the AGN-host co-evolution. which is more prominent for
          higher luminosity systems, confirming previous results.
\item We do not find any correlation between the star-formation rate (or the
          specific SFR, or the ``starburstiness'') and the X-ray absorption derived from
          high-quality {\it XMM-Newton} spectra, at any redshift or X-ray luminosity. We
          assume that this is an indication that the X-ray absorption is linked to the
          nuclear region, and the star-formation to the host.
\item Comparing the sSFR of the hosts to the characteristic sSFR of star-forming
          galaxies at the same redshift (``main sequence'') we find that the AGNs reside
          mostly in main-sequence and starburst galaxies, with the mean specific SFR
          being close the limit between main-sequence and starburst hosts. This
          reflects the AGN-starburst connection.
\item Higher X-ray luminosity AGNs (X-ray QSOs with
          $L_{\rm x}>10^{44}{\rm \,erg\,s^{-1}}$) are found in starburst hosts with average
          sSFR more than double that of the ``main sequence'' at any redshift above
          $z\approx2$. At lower redshifts ($z\approx1.5$) we find a number of QSOs
          with low sSFR values, which drive the mean starburstiness of QSOs to a value
          consistent with that of the overall AGN population.
\item We test the reliability of the colour-magnitude diagram in assessing the host
          properties, and find a significant anti-correlation between the ``redness''
          (deviation of the rest-frame colours from the line dividing red and blue
          galaxies, without any correction for AGN contribution or dust extinction), and
          the ``starburstiness'' (the sSFR divided by the ``main sequence'' sSFR at a
          given redshift).
\end{enumerate}

\begin{acknowledgements}
We acknowledge financial contribution from the agreement ASI-INAF I/009/10/00.
ER acknowledges financial support from the Marie-Curie Fellowship grant RF040294.
FJC acknowledges financial support for this work by the Spanish Ministry of Science
and Innovation through the grant AYA2010-21490-C02-01. DMA and ADM
acknowledge support from the STFC.
\end{acknowledgements}

%\clearpage


\begin{thebibliography}{}
\scriptsize{
\bibitem[Akritas \& Siebert(1996)]{Akritas1996} Akritas, M. G., \& Siebert, J. 1996, MNRAS, 278, 919
\bibitem[Alexander \& Hickox(2012)]{Alexander2012} Alexander, D. M., \& Hickox, R. C. 2012, New A Rev., 56, 93
\bibitem[Alexander et al.(2005)]{Alexander2005} Alexander, D. M., Bauer, F. E., Chapman, S. C., et al. 2005, ApJ, 632, 736
\bibitem[Appleton et al.(2004)]{Appleton2004} Appleton, P. N., Fadda, D. T., Marleau, F. R., et al. 2004, ApJS, 154, 147
\bibitem[Balestra et al.(2010)]{Balestra2010} Balestra, I., Mainieri, V., Popesso, P. et al., 2010, A\&A, 512, 12
\bibitem[Ballantyne(2008)]{Ballantyne2008} Ballantyne, D. R. 2008, ApJ, 685, 787
\bibitem[Ballantyne et al.(2006)Ballantyne, Everett \& Murray]{Ballantyne2006} Ballantyne, D. R., Everett, J. E., \& Murray, N. 2006, ApJ, 639, 740
\bibitem[Barnes \& Hernquist(1996)]{Barnes1996} Barnes, J. E., \& Hernquist, L. 1996, ApJ, 471, 115
\bibitem[Bauer et al.(2004)]{Bauer2004} Bauer, F. E., Alexander, D. M., Brandt, W. N., et al. 2004, AJ, 128, 2048
\bibitem[Bauer et al.(2002)]{Bauer2002} Bauer, F. E., Alexander, D. M., Brandt, W. N., et al. 2002, AJ, 124, 2351
\bibitem[Bell(2003)]{Bell2003} Bell, E. F. 2003, ApJ, 586, 794
\bibitem[Bolzonella et al.(2010)]{Bolzonella2010} Bolzonella, M., Kova\v{c}, K., Pozzetti, L., et al. 2010, A\&A, 524, 76
\bibitem[Bonfield et al.(2011)]{Bonfield2011} Bonfield, D. G., Jarvis, M. J., Hardcastle, M. J., et al. 2011, MNRAS, 416, 13
\bibitem[Borch et al.(2006)]{Borch2006} Borch, A., Meisenheimer, K., Bell, E. F., et al. 2006, A\&A, 453, 869
\bibitem[Bournaud et al.(2011)]{Bournaud2011} Bournaud, F., Dekel, A., Teyssier, R., et al. 2011, ApJ, 741L, 33
\bibitem[Brinchmann et al.(2004)]{Brinchmann2004} Brinchmann, J., Charlot, S., White, S. D. M., et al. 2004, MNRAS, 351, 1151
\bibitem[Brusa et al.(2009)]{Brusa2009} Brusa, M., Fiore, F., Santini, P., et al. 2009, A\&A, 507, 1277
\bibitem[Bruzual \& Charlot(2003)]{Bruzual2003} Bruzual, G., \& Charlot, S. 2003, MNRAS, 344, 1000
\bibitem[Buckley \& James(1979)]{Buckley1979} Buckley, J., \& James, I. 1979, Biometrika, 66, 429
\bibitem[Calzetti et al.(2000)]{Calzetti2000} Calzetti, D., Armus, L., Bohlin, R. C., et al. 2000, ApJ, 533, 682
%\bibitem[Caputi et al.(2007)]{Caputi2007} Caputi, K. I., Lagache, G., Yan, L., et al. 2007, ApJ, 660, 97
\bibitem[Cardamone et al.(2010a)]{Cardamone2010a} Cardamone, C. N., van Dokkum, P. G., Urry, C. M. et al., 2010, ApJS, 189, 270
\bibitem[Cardamone et al.(2010b)]{Cardamone2010b} Cardamone, C. N., Urry, M., Schawinski, K., et al. 2010, ApJ, 721L, 38
\bibitem[Casey et al.(2011)]{Casey2011} Casey, C. M., Chapman, S. C., Smail, I., et al. 2011, MNRAS, 411, 2739
\bibitem[Chabrier(2003)]{Chabrier2003} Chabrier, G. 2003, ApJ, 586L, 133
\bibitem[Chapman et al.(2005)]{Chapman2005} Chapman, S. C., Blain, A. W., Smail, I., \& Ivison, R. J. 2005, ApJ, 622, 772
\bibitem[Chary \& Elbaz(2001)]{Chary2001} Chary, R., \& Elbaz, D. 2001, ApJ, 556, 562
\bibitem[Cisternas et al.(2011)]{Cisternas2011} Cisternas, M., Jahnke, K., Inskip, K. J., et al. 2011, ApJ, 726, 57
\bibitem[Comastri et al.(2011)]{Comastri2011} Comastri, A., Ranalli, P., Iwasawa, K., et al 2011, A\&A, 526L, 9
\bibitem[Condon(1992)]{Condon1992} Condon, J. J. 1992, ARA\&A, 30, 575
\bibitem[Cooper et al.(2011)]{Cooper2011} Cooper, M. C., Yan, R., Dickinson, M., et al. 2011, MNRAS, submitted {\texttt [arXiv: astro-ph/1112.0.12v1]}
\bibitem[Daddi et al.(2010)]{Daddi2010} Daddi, E., Bournaud, F., Walter, F., et al. 2010, ApJ, 713, 686
\bibitem[Daddi et al.(2009)]{Daddi2009} Daddi, E., Dannerbauer, H., Stern, D., et al. 2009, ApJ, 694, 1517
\bibitem[Daddi et al.(2007)]{Daddi2007} Daddi, E., Dickinson, M., Morrison, G., et al. 2007, ApJ, 670, 156
\bibitem[Damen et al.(2011)]{Damen2011} Damen, M., Labb\'{e}, I., van Dokkum, P. G., et al. 2011, ApJ, 727, 1
%\bibitem[Dasyra et al.(2008)]{Dasyra2008} Dasyra, K. M., Yan, L., Helou, G., 2008, ApJ, 680, 232
\bibitem[Di Matteo et al.(2008)]{DiMatteo2008} Di Matteo, T., Colberg, J., Springel, V., Hernquist, L., \& Sijacki, D. 2008, ApJ, 676, 33
\bibitem[Di Matteo et al.(2005)Di Matteo, Springel \& Hernquist]{DiMatteo2005} Di Matteo, T., Springel, V., \& Hernquist, L. 2005, Nat, 433, 604
\bibitem[Diamond-Stanic \& Rieke(2012)]{DiamondStanic2012} Diamond-Stanic, A. M., \& Rieke, G. H. 2012, ApJ, 746, 168
\bibitem[Dunne et al.(2009)]{Dunne2009} Dunne, L., Ivison, R. J., Maddox, S., et al. 2009, MNRAS, 394, 3
\bibitem[Elbaz et al.(2011)]{Elbaz2011} Elbaz, D., Dickinson, M., Hwang, H. S., et al. 2011, A\&A, 533, 119
\bibitem[Elbaz et al.(2009)]{Elbaz2009} Elbaz, D., Jahnke, K., Pantin, E., Le Borgne, D., \& Letawe, G. 2009, A\&A, 507, 1359
\bibitem[Elbaz et al.(2007)]{Elbaz2007} Elbaz, D., Daddi, E., Le Borgne, D., et al. 2007, A\&A, 468, 33
\bibitem[Elitzur \& Shlosman(2006)]{Elitzur2006} Elitzur, M., \& Shlosman, I. 2006, ApJ, 648L, 101
\bibitem[Elvis et al.(1994)]{Elvis1994} Elvis, M., Wilkes, B. J., McDowell, J. C., et al. 1994, ApJS, 95, 1
\bibitem[Feain et al.(2007)]{Feain2007} Feain, I. J., Papadopoulos, P. P., Ekers, R. D., \& Middelberg, E. 2007, ApJ, 662, 872
\bibitem[Feigelson \& Nelson(1985)]{Feigelson1985} Feigelson, E. D., \& Nelson, P. I. 1985, ApJ, 293, 192
\bibitem[Ferrarese \& Merritt (2000)]{Ferrarese2000} Ferrarese, L., \& Merritt, D. 2000, ApJ, 539, L9
\bibitem[F\"{o}rster Schreiber et al.(2009)]{ForsterSchreiber2009} F\"{o}rster Schreiber, N. M., Genzel, R., Bouch\'{e}, N., et al. 2009, ApJ, 706, 1364
\bibitem[Gaibler et al.(2011)]{Gaibler2011} Gaibler, V., Khochfar, S., Krause, M., \& Silk, J. 2011, MNRAS, in press {\texttt [arXiv: astro-ph/1111.4478v1]}
\bibitem[Gawiser et al.(2006)]{Gawiser2006} Gawiser, E., van Dokkum, P. G., Herrera, D., et al. 2006, ApJS, 162, 1
\bibitem[Georgakakis \& Nandra(2011)]{Georgakakis2011} Georgakakis, A., \& Nandra, K. 2011, MNRAS, 414, 992
\bibitem[Georgakakis et al.(2008)]{Georgakakis2008} Georgakakis, A., Nandra, K., Yan, R., et al. 2008, MNRAS, 385, 2049
\bibitem[Georgakakis et al.(2006)]{Georgakakis2006} Georgakakis, A., Georgantopoulos, I., Akylas, A., Zezas, A., \& Tzanavaris, P. 2006, ApJ, 641L, 101
\bibitem[Georgakakis et al.(2004)]{Georgakakis2004} Georgakakis, A., Hopkins, A. M., Afonso, J., et al. 2004, MNRAS, 354, 127
\bibitem[Georgantopoulos et al.(2011a)]{Georgantopoulos2011a} Georgantopoulos, I., Dasyra, K. M., Rovilos, E., et al. 2011, A\&A, 531, 116
\bibitem[Georgantopoulos et al.(2011b)]{Georgantopoulos2011b} Georgantopoulos, I., Rovilos, E., Akylas, A., et al. 2011, A\&A, 534, 23
\bibitem[Giroletti \& Panessa(2009)]{Giroletti2009} Giroletti, M., \& Panessa, F. 2009, ApJ, 706L, 260
\bibitem[Gonz\'{a}lez et al.(2010)]{Gonzalez2010} Gonz\'{a}lez, V., Labb\'{e}, I., Bouwens, R. J., et al. 2010, ApJ, 713, 115
\bibitem[Griffin et al.(2010)]{Griffin2010} Griffin, M.J., Abergel, A., Abreu, A., et al. 2010, A\&A, 518, L3
\bibitem[Grogin et al.(2005)]{Grogin2005} Grogin, N. A., Conselice, C. J., Chatzichristou, E., et al. 2005, ApJ, 627L, 97
\bibitem[G\"{u}ltekin et al.(2009)]{Gultekin2009} G\"{u}ltekin, K., Richstone, D. O., Gebhardt, K., et al. 2009, ApJ, 698, 198
\bibitem[Hainline et al.(2011)]{Hainline2011} Hainline, L. J., Blain, A. W., Smail, I., et al. 2011, ApJ, 740, 96
\bibitem[Hasinger(2008)]{Hasinger2008} Hasinger, G. 2008, A\&A, 490, 905
\bibitem[Hatziminaoglou et al.(2010)]{Hatziminaoglou2010} Hatziminaoglou, E., Omont, A., Stevens, J. A., et al. 2010, A\&A, 518L, 33
\bibitem[Hickox et al.(2009)]{Hickox2009} Hickox, R. C., Jones, C., Forman, W. R., et al. 2009, ApJ, 696, 891
\bibitem[Hopkins \& Quataert(2010)]{Hopkins2010a} Hopkins, P. F., \& Quataert, E. 2010, MNRAS, 407, 1529
%\bibitem[Hopkins et al.(2010)]{Hopkins2010b} Hopkins, P. F., Younger, J. D., Hayward, C. C., Narayanan, D., \& Hernquist, L. 2010, MNRAS, 402, 1693
\bibitem[Hopkins et al.(2006)]{Hopkins2006} Hopkins, P. F., Hernquist, L., Cox, T. J., et al. 2006, ApJS, 163, 1
%\bibitem[Hwang et al.(2012)]{Hwang2012} Hwang, H. S., Park, C., Elbaz, D., \& Choi, Y.-Y. 2012, A\&A, 538, 15
\bibitem[Isobe et al.(1986)Isobe, Feigelson \& Nelson]{Isobe1986} Isobe, T., Feigelson, \& E. D., Nelson, P. I. 1986, ApJ, 306, 490
\bibitem[Ivison et al.(2010)]{Ivison2010} Ivison, R. J., Magnelli, B., Ibar, E., et al. 2010, A\&A, 518L, 31
\bibitem[Karim et al.(2011)]{Karim2011} Karim, A., Schinnerer, E., Mart\'{i}nez-Sansigre, A., et al. 2011, ApJ, 730, 61
\bibitem[Kartaltepe et al.(2012)]{Kartaltepe2012} Kartaltepe, J. S., Dickinson, M., Alexander, D. M., et al. 2012, ApJ, submitted {\texttt [arXiv: astro-ph/1110.4057v2]}
\bibitem[Kauffmann et al.(2003)]{Kauffmann2003} Kauffmann, G., Heckman, T. M., Tremonti, C., et al. 2003, MNRAS, 346, 1055
\bibitem[Kellermann et al.(2008)]{Kellermann2008} Kellermann, K. I., Fomalont, E. B., Mainieri, V., et al. 2008, ApJS, 179, 71
\bibitem[Kellermann et al.(1989)]{Kellermann1989} Kellermann, K. I., Sramek, R., Schmidt, M., Shaffer, D. B., \& Green, R. 1989, AJ, 98, 1195
\bibitem[Kelly et al.(2007)]{Kelly2007} Kelly, B. C., Bechtold, J., Siemiginowska, A., Aldcroft, T., \& Sobolewska, M. 2007, ApJ, 657, 116
\bibitem[Kennicutt \& Evans(2012)]{Kennicutt2012} Kennicutt, R. C., \& Evans, N. J. 2012, ARA\&A, in press {\texttt [arXiv: astro-ph/1204.3552v1]}
\bibitem[Kennicutt(1998a)]{Kennicutt1998a} Kennicutt, R. C., Jr. 1998, ARA\&A, 36, 189K
%\bibitem[Kennicutt(1998b)]{Kennicutt1998b} Kennicutt, R. C., Jr. 1998, ApJ, 498, 541
\bibitem[King(2005)]{King2005} King, A. 2005, ApJ, 635L, 121
\bibitem[Kocevski et al.(2012)]{Kocevski2012} Kocevski, D. D., Faber, S. M., Mozena, M., et al. 2012, ApJ, 744, 148
\bibitem[Kormendy \& Kennicutt(2004)]{Kormendy2004} Kormendy, J., \& Kennicutt, R. C., Jr. 2004, ARA\&A, 42, 603
\bibitem[Kriek et al.(2008)]{Kriek2008} Kriek, M., van Dokkum, P. G., Franx, M., et al. 2008, ApJ, 677, 219
\bibitem[Kroupa(2001)]{Kroupa2001} Kroupa, P. 2001, MNRAS, 322, 231
\bibitem[Laird et al.(2010)]{Laird2010} Laird, E. S., Nandra, K., Pope, A., \& Scott, D. 2010, MNRAS, 401, 2763
\bibitem[LaValley et al.(1992)LaValley, Isobe \& Feigelson]{LaValley1992} LaValley, M. P., Isobe, T., \& Feigelson, E. D. 1992, BAAS, 24, 839
\bibitem[Le F\`{e}vre et al.(2005)]{LeFevre2005} Le F\`{e}vre, O., Vettolani, G., Garilli, B., et al. 2005, A\&A, 439, 845
\bibitem[Le F\`{e}vre et al.(2004)]{LeFevre2004} Le F\`{e}vre, O., Vettolani, G., Paltani, S., et al. 2004, A\&A, 428, 1043
\bibitem[Lehmer et al.(2005)]{Lehmer2005} Lehmer, B. D., Brandt, W. N., Alexander, D. M., et al. 2005, ApJS, 161, 21
\bibitem[Luo et al.(2010)]{Luo2010} Luo, B., Brandt, W. N., Xue, Y. Q., et al. 2010, ApJS, 187, 560
\bibitem[Lusso et al.(2012)]{Lusso2012} Lusso, E., Comastri, A., Simmons, B. D., et al. 2012, MNRAS, in press {\texttt [arXiv: astro-ph/1206.2642]}
\bibitem[Lusso et al.(2011)]{Lusso2011} Lusso, E., Comastri, A., Vignali, C., et al. 2011, A\&A, 534, 110
\bibitem[Lutz et al.(2011)]{Lutz2011} Lutz, D., Poglitsch, A., Altieri, B., et al. 2011, A\&A, 532, 90
\bibitem[Lutz et al.(2010)]{Lutz2010} Lutz, D., Mainieri, V., Rafferty, D., et al. 2010, ApJ, 712, 1287
\bibitem[Magdis et al.(2011)]{Magdis2011} Magdis, G. E., Elbaz, D., Dickinson, M., et al. 2011, A\&A, 534, 15
\bibitem[Magdis et al.(2010)]{Magdis2010} Magdis, G. E., Rigopoulou, D., Huang, J.-S., \& Fazio, G. G. 2010, MNRAS, 401, 1521
\bibitem[Magorrian et al.(1998)]{Magorrian1998} Magorrian, J., Tremaine, S., Richstone, D., et al. 1998, AJ, 115, 2285
\bibitem[Magnelli et al.(2011)]{Magnelli2011} Magnelli, B., Elbaz, D., Chary, R. R., et al. 2011, A\&A, 528, 35
\bibitem[Magnelli et al.(2009)]{Magnelli2009} Magnelli, B., Elbaz, D., Chary, R. R., et al. 2009, A\&A, 496, 57
\bibitem[Mart\'{i}nez-Sansigre et al.(2009)]{MartinezSansigre2009} Mart\'{i}nez-Sansigre, A., Karim, A., Schinnerer, E., et al. 2009, ApJ, 706, 184
%\bibitem[Melbourne et al.(2008)]{Melbourne2008} Melbourne, J., Ammons, M., Wright, S. A., et al. 2008, AJ, 135, 1207
\bibitem[Merloni \& Heinz(2008)]{Merloni2008} Merloni, A., \& Heinz, S. 2008, MNRAS, 388, 1011
\bibitem[Micha\l owski et al.(2012)]{Michalowski2012} Micha\l owski, M. J., Dunlop, J. S., Cirasuolo, M., et al. 2012, A\&A, 541, 85
\bibitem[Middelberg et al.(2011)]{Middelberg2011} Middelberg, E., Deller, A., Morgan, J., et al. 2011, A\&A, 526, 74
\bibitem[Mignoli et al.(2005)]{Mignoli2005} Mignoli, M., Cimatti, A., Zamorani, G., et al. 2005, A\&A, 437, 883
\bibitem[Miller et al.(2008)]{Miller2008} Miller, N. A., Fomalont, E. B., Kellermann, K. I., et al. 2008, ApJS, 179, 114
\bibitem[Miller et al.(1990)Miller, Peacock \& Mead]{Miller1990} Miller, L., Peacock, J. A., \& Mead, A. R. G. 1990, MNRAS, 244, 207
\bibitem[Mullaney et al.(2012a)]{Mullaney2012} Mullaney, J. R., Pannella, M., Daddi, E., et al. 2012a, MNRAS, 419, 95
\bibitem[Mullaney et al.(2012b)]{Mullaney2012b} Mullaney, J. R., Daddi, E., Bethermin, M., et al. 2012b, ApJ, 753L, 30
\bibitem[Mullaney et al.(2011)]{Mullaney2011} Mullaney, J. R., Alexander, D. M., Goulding, A. D., \& Hickox, R. C. 2011, MNRAS, 414, 1082
\bibitem[Mullaney et al.(2010)]{Mullaney2010} Mullaney, J. R., Alexander, D. M., Huynh, M., Goulding, A. D., \& Frayer, D. 2010, MNRAS, 401, 995
\bibitem[Murphy et al.(2011)]{Murphy2011} Murphy, E. J., Condon, J. J., Schinnerer, E., et al. 2011, ApJ, 737, 67
\bibitem[Nandra et al.(2007)]{Nandra2007} Nandra, K., Georgakakis, A., Willmer, C. N. A., et al. 2007, ApJ, 660L, 11
\bibitem[Nenkova et al.(2008)]{Nenkova2008} Nenkova, M., Sirocky, M. M., Nikutta, R., Ivezi\'{c}, \v{Z}., \& Elitzur, M. 2008, ApJ, 685, 160
\bibitem[Netzer(2009)]{Netzer2009} Netzer, H. 2009, MNRAS, 399, 1907
\bibitem[Norris et al.(2006)]{Norris2006} Norris, R. P., Afonso, J., Appleton, P. N., et al. 2006, AJ, 132, 2409
\bibitem[Page et al.(2012)]{Page2012} Page, M. J., Symeonidis, M., Vieira, J., D., et al. 2012, Nature, 485, 213
\bibitem[Page et al.(2004)]{Page2004} Page, M. J., Stevens, J. A., Ivison, R. J., \& Carrera, F. J. 2004, ApJ, 611L, 85
\bibitem[Page et al.(2001)]{Page2001} Page, M. J., Stevens, J. A., Mittaz, J. P. D., \& Carrera, F. J. 2001, Sci, 294, 2516
\bibitem[Pannella et al.(2009)]{Pannella2009} Pannella, M., Carilli, C. L., Daddi, E., et al. 2009, ApJ, 698L, 116
\bibitem[Papovich et al.(2001)Papovich, Dickinson \& Ferguson]{Papovich2001} Papovich, C., Dickinson, M., \& Ferguson, H. C. 2001, ApJ, 559, 620
\bibitem[Peng et al.(2010)]{Peng2010} Peng, Y.-j., Lilly, S. J., Kova\v{c}, K., et al. 2010, ApJ, 721, 193
\bibitem[Pierce et al.(2010)]{Pierce2010} Pierce, C. M., Lotz, J. M., Primack, J. R., et al. 2010, MNRAS, 405, 718
\bibitem[Padovani et al.(2011)]{Padovani2011} Padovani, P., Miller, N., Kellermann, K. I., et al. 2011, ApJ, 740, 20
\bibitem[Poglitsch et al.(2010)]{Poglitsch2010} Poglitsch, A., Waelkens, C., Geis, N., et al. 2010, A\&A, 518, L2
\bibitem[Pozzetti et al.(2010)]{Pozzetti2010} Pozzetti, L., Bolzonella, M., Zucca, E., et al. 2010, A\&A, 523, 13
\bibitem[Rafferty et al.(2011)]{Rafferty2011} Rafferty, D. A., Brandt, W. N., Alexander, D. M., et al. 2011, ApJ, 742, 3
\bibitem[Ranalli et al.(2003)Ranalli, Comastri \& Setti]{Ranalli2003} Ranalli, P., Comastri, A., \& Setti, G. 2003, A\&A, 399, 39
\bibitem[Ravikumar et al.(2007)]{Ravikumar2007} Ravikumar, C. D., Puech, M., Flores, H., et al. 2007, A\&A, 465, 1099
\bibitem[Rodighiero et al.(2011)]{Rodighiero2011} Rodighiero, G., Daddi, E., Baronchelli, I., et al. 2011, ApJ, 739L, 40R
\bibitem[Rosario et al.(2012)]{Rosario2012} Rosario, D. J., Santini, P., Lutz, D., et al. 2012, A\&A, submitted {\texttt [arXiv: astro-ph/1203.6069v1]}
\bibitem[Rovilos \& Georgantopoulos(2007)]{Rovilos2007} Rovilos, E., \& Georgantopoulos, I. 2007, A\&A, 475, 115
\bibitem[Rovilos et al.(2011)]{Rovilos2011} Rovilos, E., Fotopoulou, S., Salvato, M., et al. 2011, A\&A, 529, 135
\bibitem[Rovilos et al.(2007)]{Rovilos2007b} Rovilos, E., Georgakakis, A., Georgantopoulos, I., et al. 2007, A\&A, 466, 119
\bibitem[Salim et al.(2007)]{Salim2007} Salim, S., Rich, R. M., Charlot, S., et al. 2007, ApJS, 173, 267
%\bibitem[Salpeter(1955)]{Salpeter1955} Salpeter, E. E. 1955, ApJ, 121, 161
%\bibitem[Sanders et al.(1988)]{Sanders1988} Sanders, D. B., Soifer, B. T., Elias, J. H., Neugebauer, G., \& Matthews, K. 1988, ApJ, 328L, 35
\bibitem[Santini et al.(2012)]{Santini2012} Santini, P., Rosario, D., Shao, L., et al. 2012, A\&A, 540, 109
\bibitem[Scott et al.(2010)]{Scott2010} Scott, K. S., Yun, M. S., Wilson, G. W., et al. 2010, MNRAS, 405, 2260
\bibitem[Serjeant \& Hatziminaoglou(2009)]{Serjeant2009} Serjeant, S., \& Hatziminaoglou, E. 2009, MNRAS, 397, 265
\bibitem[Schartmann et al.(2009)]{Schartmann2009} Schartmann, M., Meisenheimer, K., Klahr, H., et al. 2009, MNRAS, 393, 759
\bibitem[Seymour et al.(2011)]{Seymour2011} Seymour, N., Symeonidis, M., Page, M. J., et al. 2011, MNRAS, 413, 1777
\bibitem[Shao et al.(2010)]{Shao2010} Shao, L., Lutz, D., Nordon, R., et al. 2010, A\&A, 518L, 26
\bibitem[Shapley et al.(2001)]{Shapley2001} Shapley, A. E., Steidel, C. C., Adelberger, K. L., et al. 2001, ApJ, 562, 95
\bibitem[Silva et al.(2004)Silva, Maiolino \& Granato]{Silva2004} Silva, L., Maiolino, R., \& Granato, G. L. 2004, MNRAS, 355, 973
\bibitem[Silverman et al.(2010)]{Silverman2010} Silverman, J. D., Mainieri, V., Salvato, M., et al. 2010, ApJS, 191, 124
\bibitem[Silverman et al.(2009)]{Silverman2009} Silverman, J. D., Lamareille, F., Maier, C., et al. 2009, ApJ, 696, 396
\bibitem[Silverman et al.(2008)]{Silverman2008} Silverman, J. D., Mainieri, V., Lehmer, B. D., et al. 2008, ApJ, 675, 1025
\bibitem[Springel et al.(2005)Springel, Di Matteo \& Hernquist]{Springel2005} Springel, V., Di Matteo, T., \& Hernquist, L. 2005, ApJ, 620L, 79
\bibitem[Stark et al.(2009)]{Stark2009} Stark, D. P., Ellis, R. S., Bunker, A., et al. 2009, ApJ, 697, 1493
\bibitem[Stocke et al.(1991)]{Stocke1991} Stocke, J. T., Morris, S. L., Gioia, I. M., et al. 1991, ApJS, 76, 813
\bibitem[Sutherland \& Saunders(1992)]{Sutherland1992} Sutherland, W., \& Saunders, W. 1992, MNRAS, 259, 413
\bibitem[Szokoly et al.(2004)]{Szokoly2004} Szokoly, G. P., Bergeron, J., Hasinger, G., et al. 2004, ApJS, 155, 271
\bibitem[Taylor et al.(2009)]{Taylor2009} Taylor, E. N., Franx, M., van Dokkum, P. G., et al. 2009, ApJS, 183, 295
\bibitem[Thompson et al.(2009)]{Thompson2009} Thompson, G. D., Levenson, N. A., Uddin, S. A., \& Sirocky, M. M. 2009, ApJ, 697, 182
\bibitem[Treister et al.(2009)]{Treister2009} Treister, E., Cardamone, C. N., Schawinski, K., et al. 2009, ApJ, 706, 535
\bibitem[Trichas et al.(2012)]{Trichas2012} Trichas, M., Green, P. J., Silverman, J. D., et al. 2012, ApJS, 200, 17
\bibitem[Trichas et al.(2009)]{Trichas2009} Trichas, M., Georgakakis, A., Rowan-Robinson, M., et al. 2009, MNRAS, 399, 663
\bibitem[Tzanavaris et al.(2006)Tzanavaris, Georgantopoulos \& Georgakakis]{Tzanavaris2006} Tzanavaris, P., Georgantopoulos, I., \& Georgakakis, A. 2006, A\&A, 454, 447
\bibitem[van der Wel et al.(2005)]{vanderWel2005} van der Wel, A., Franx, M., van Dokkum, P. G., et al. 2005, ApJ, 631, 145
\bibitem[Vanzella et al.(2008)]{Vanzella2008} Vanzella, E., Cristiani, S., Dickinson, M., et al. 2008, A\&A, 478, 83
\bibitem[Virani et al.(2006)]{Virani2006} Virani, S. N., Treister, E., Urry, C. M., \& Gawiser, E. 2006, AJ, 131, 2373
\bibitem[Wei\ss\ et al.(2009)]{Weiss2009} Wei\ss, A., Kov\'{a}cs, A., Coppin, K., et al. 2009, ApJ, 707, 1201
\bibitem[White et al.(2000)]{White2000} White, R. L., Becker, R. H., Gregg, M. D., et al. 2000, ApJS, 126, 133
\bibitem[Wilman et al.(2010)]{Wilman2010} Wilman, R. J., Jarvis, M. J., Mauch, T., Rawlings, S., \& Hickey, S. 2010, MNRAS, 405, 447
\bibitem[Xue et al.(2011)]{Xue2011} Xue, Y. Q., Luo, B., Brandt, W. N., et al. 2011, ApJS, 195, 10
\bibitem[Xue et al.(2010)]{Xue2010} Xue, Y. Q.,  Brandt, W. N., Luo, B., et al. 2010, ApJ, 720, 368
}
\end{thebibliography}
\end{document}